%% file: Kolmogorov-2023-replace.tex
\documentclass[11pt,a4paper]{article}
\usepackage{hyperref}
\usepackage{tikz-cd}
\usepackage[utf8]{inputenc}
\usepackage{graphicx}  \usepackage{mathptmx}
\usepackage{amsmath, amssymb}
\numberwithin{equation}{section}
\usepackage{gensymb}

\input{Bohrificationpreamble.tex}

\newtheorem{Definition}{Definition}[section]

\newtheorem{Theorem}[Definition]{Theorem}

\newcommand{\ML}{Martin-L\"{o}f}
\newcommand{\ut}{\underline{2}}
\newcommand{\usg}{\underline{\sigma}}
\topmargin = - 1 cm \textheight = 23 cm \textwidth = 15 cm
\usepackage[symbol]{footmisc}
\renewcommand{\thefootnote}{\fnsymbol{footnote}}
\begin{document}
\pagenumbering{arabic} \setlength{\unitlength}{1cm}\cleardoublepage
\date\nodate
\begin{center}
\begin{Huge}
{\bf Typical =  random}\footnote[1]{This paper originated in a talk at the ``Kolmogorov Workshop'' in Geneva, February 16, 2023. I thank Alexei Grinbaum for the invitation, as well as various participants for questions during and after the talk. My  presentation of \ML\ randomness below benefited from discussions with Cristian Calude.  I also thank Chris Porter and Sylvia Wenmackers for helpful correspondence.  The recent volume \emph{Algorithmic Randomness: Progress and Prospects},  eds.\ J.Y. Franklin and C.P. Porter (CUP, 2020), greatly faciliated this work.
}
\end{Huge}
\bigskip\bigskip

\begin{Large}
 Klaas Landsman\vspace{5mm}
 \end{Large}
 
 \begin{large}
Institute for Mathematics, Astrophysics, and  Particle Physics\\ \vspace{1mm}
 Radboud Center for Natural Philosophy  \\ \vspace{1mm}
Radboud University, Nijmegen, The Netherlands\\ \vspace{1mm}
\texttt{landsman@math.ru.nl}
\end{large}\bigskip\bigskip

\fbox{\emph{Dedicated to the memory of Marinus Winnink (1936--2023)}}
\bigskip

 \begin{abstract} 
\noindent This expository paper advocates an approach to physics in which ``typicality" is identified with a suitable form of algorithmic randomness. To this end various theorems from mathematics and physics are reviewed. Their original versions state that some property $F(x)$ holds \emph{for $P$-almost all  $x\in X$}, where $P$ is a  probability measure on some space $X$. Their more refined (and typically more recent) formulations show that $F(x)$ holds \emph{for all $P$-random  $x\in X$}.  The computational  notion of $P$-randomness used here generalizes the one introduced by \ML\ in 1966 in a way now standard in algorithmic randomness. Examples come from probability theory, analysis, dynamical systems/ergodic theory, statistical mechanics, and quantum mechanics (especially hidden variable theories). An underlying philosophical theme, inherited from von Mises and Kolmogorov,  is the interplay between probability and randomness, especially: which comes first?
 \end{abstract}\end{center}
\tableofcontents
\thispagestyle{empty}
\renewcommand{\thefootnote}{\arabic{footnote}}
\newpage \setcounter{footnote}{0}
\section{Introduction}
The introduction of probability in statistical mechanics in the 19th century by Maxwell and  Boltzmann immediately raised questions about both the meaning of this concept by itself and its relationship to  randomness and entropy (Brush, 1976;  Sklar, 1993;  von Plato, 1994;  Uffink, 2007, 2022). Roughly speaking, both  initially felt that probabilities in statistical mechanics were dynamically generated by particle trajectories, a view which led to ergodic theory. But subsequently  Boltzmann (1877) introduced his  counting arguments as a new start; these led to his famous formula for the entropy $S=k \log W$ on his gravestone in Vienna, i.e.,
\begin{center}
\emph{probability first, entropy second}.
\end{center}
 This was turned on its head by Einstein (1909), who had rediscovered much of statistical mechanics by himself in his early work, always stressing the role of fluctuations. He expressed the probability of energy fluctuations in terms of entropy seen as a primary concept. This suggests:
 \begin{center}
\emph{entropy first,  probability second}.
\end{center}
  From the modern point of view of large deviation theory
 (Lanford, 1973; A.\ \ML, 1979--the older brother of P.\ \ML--; Ellis, 1985), what happens is that for finite $N$ some stochastic process $(X_N)$  fluctuates around its limiting value $\ovl{X}$ as $N\raw\infty$ (if it has one), and, under favorable circumstances that often obtain in statistical mechanics, the ``large'' i.e.\ $O(1)$ fluctuations (as opposed to the $O(1/\sqrt{N})$ fluctuations, which are described by the central limit theorem, cf.\ McKean, 2014) can be computed via an entropy function  $S(x)$ 
 whose argument $x$  lies in the (common) codomain $\mathcal{X}$ of $X_N:\Om\raw \mathcal{X}$. Since the domain  $\Om$ of $X_N$ carries a probability measure to begin with, it  seems an illusion that entropy could be defined without \emph{some} prior notion of probability. 
 
 Similar question may be asked about the connection between probability and randomness (and, closing the triangle, of course also about the relationship between randomness and entropy).
 First, 
 in his influential (but flawed) work on the foundations of probability, von Mises (1919, 1936) initially defined randomness through  a \emph{Kollektiv} (which, with hindsight, was a precursor to a random sequence). From this, he extracted a notion of probability via asymptotic relative frequencies. See also  van Lambalgen (1987, 1996), von Plato (1994), and Porter (2012).
 Von Plato (1994, p.\ 190) writes that `He [von Mises] was naturally aware of the earlier attempts of Einstein and others at founding statistical physics on classical dynamics' and  justifies this view in his \S 6.3. 
 Thus:
  \begin{center}
\emph{randomness first,  probability second}.
\end{center}
Kolmogorov (1933), on the other hand, (impeccably) defined \emph{probability first} (via measure theory), \emph{in terms of which} he hoped to understand randomness. In other words, his (initial) philosophy was:
  \begin{center}
\emph{probability first,  randomness second}.
\end{center}
Having realized that this was impossible, thirty years later Kolmogorov arrived at the  concept  of  randomness  named after him, using tools from computer and information science that actually had  roots in  in the work of von Mises (as well as of Turing, Shannon, and others). See van Lambalgen (1987), Cover et al. (1989), von Plato (1994),   Li \& Vit\'{a}nyi (2008), and Porter (2014).  So:
  \begin{center}
\emph{Kolmogorov randomness first, measure-theoretic probability second}.
\end{center}
\noindent But I will argue that even Kolmogorov randomness  seems to rely on some prior concept of probability, see \S\ref{AR} and in particular the discussion surrounding Theorem \ref{LG}; and this is obviously the case for \ML-randomness, both in its original form for binary sequences (which is essentially equivalent to 
Kolmogorov randomness as extended from finite strings to infinite sequences, see Theorem \ref{Calthm}) and in its generalizations (see \S\ref{AR}). So I will defend the view that after all we have
  \begin{center}
\emph{some prior probability measure first, \ML\ randomness second}.
\end{center}
 In any case, there isn't a single concept of randomness (Landsman, 2020), not even within the algorithmic setting (Porter, 2021); although the above slogan probably  applies to most of them.

Motivated by the above discussion and its potential applications to physics, the aim of this paper is to review the interplay between probability, (algorithmic) randomness, and  entropy via examples from probability itself, analysis, dynamical systems and (Boltzmann-style) statistical mechanics, and quantum mechanics.  Some basic relations are explained in the next \S \ref{BEP}.  In \S\ref{AR} I review algorithmic randomness beyond binary sequences. Section \ref{PP} introduces some key  ``intuition pumps": these are results in which `for $P$-almost every $x$: $\Phi(x)$' in some ``classical'' result can be replaced by `for all $P$-random $x$: $\Phi(x)$' in an ``effective'' counterpart thereof; this replacement may even be seen as the essence of algorithmic randomness. 
In section \ref{SM} I apply this idea to statistical mechanics, and close in 
\S\ref{QM} with some brief comments on quantum mechanics. The paper closes with a brief summary.
 \section{Some background on entropy and probability}\label{BEP}
Consider  the following diagram, which connects and illustrates the main examples in this paper.
\begin{figure}
\centering
 \includegraphics[width=0.95\textwidth]{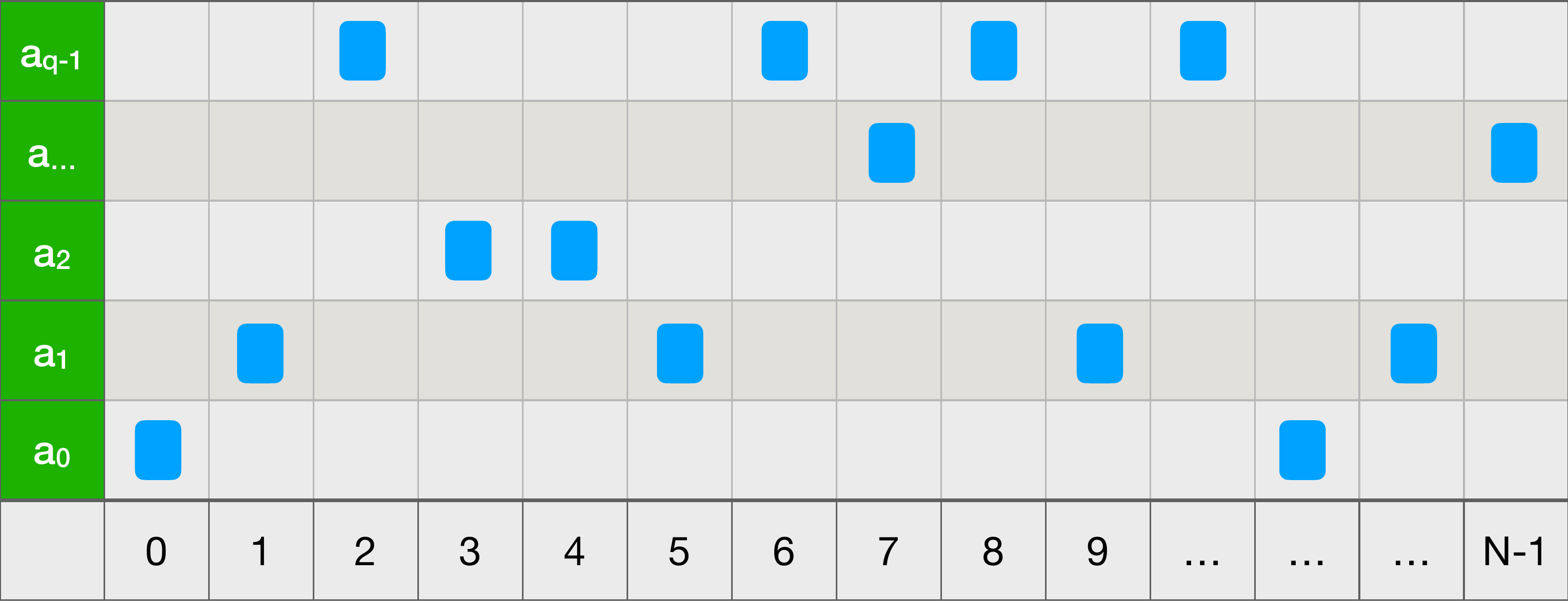}
\caption{Sample configuration on $N$ objects each of which can be in $q$ different states}
\end{figure}  
\noindent Here $N\in\N$ is meant to be some \emph{large} natural number, whereas $q\in\{2,3,\ldots\}$, the cardinality of 
\beq
A=\{a_0, \ldots, a_{q-1}\},
\eeq
 could be anything (finite),  but is \emph{small} ($q=2$) in the already interesting case of binary strings.
 In what follows, $A^N$ is the set of all functions $\sg:N\raw A$, where 
 \beq
 N=\{0, 1, \ldots, N-1\},
 \eeq as usual in set theory. Such a function is also called a \emph{string} over $A$, having length 
 \beq
 \ell(\sg)\equiv |\sg|=N.\eeq 
We write either $\sg(n)$ or $\sg_n$ for its value at $n\in N$, and may write $\sg$ as $\sg_0\sg_1\cdots \sg_{N-1}$. In particular, if $A=2=\{0,1\}$, then $\sg$ is a binary string. I write 
\beq
A^*\equiv A^{<\om}=\bigcup_{N\in\N}A^N,
\eeq so that 
 $2^*=\bigcup_{N\in\N}2^N$ is the set of all binary strings. Thus a (binary) \emph{string} $\sg$ is \emph{finite}, whereas a (binary) \emph{sequence} $s$ is \emph{infinite}. The set of all binary sequences is denoted by $2^{\om}$, and likewise $A^{\om}$ consists of all functions $s:\N\raw A$. For $s\in A^{\om}$, I write $s_{|N}$ for $s_0s_1\cdots s_{N-1}\in A^N$, to be sharply distinguished from $s_n\equiv s(n)\in A$. Using this notation, I now review various ways of looking at the above diagram. 
Especially in the first two items below it is hard to avoid  overlap with e.g.\ Georgii (2003) and Gr\"{u}nwald \&  Vit\'{a}nyi (2003), which I recommend for further information.
\begin{itemize}
\item In \emph{statistical mechanics} as developed by
Boltzmann (1877), and more generally in what one might call ``Boltzmann-style statistical mechanics'', which is  based on typicality arguments (Bricmont, 2022), $N$ is the number of (distinguishable) particles under consideration, and  $A$ could be a finite set of single-particle energy levels. More generally, $a\in A$ is some property each particle may separately have, such as its location in cell $X_a$ relative to some partition
 \begin{equation}
X=\bigsqcup_{a\in A} X_{a} \label{Xdu}
\end{equation}
of the \emph{single-particle} phase space or configuration space $X$ accessible to each particle.
Here $X_a\subset X$ and different $X_a$ are disjoint, which fact is expressed by the symbol $\bigsqcup$ in \er{Xdu}. One might replace $\bigsqcup$ by $\bigcup$ as long as one knows that the subsets $X_a$ are mutually disjoint (and measurable as appropriate).
 The \emph{microstate} $\sg\in A^N$ is a function $\sg:\{0, 1, \ldots, N-1\}\raw A$, written $n\mapsto \sg(n)$ or $n\mapsto \sg_n$, that specifies which property (among the possibilities in $A$) each particle has. Thus  also spin chains fall under this formalism, where $\sg_n\in A$ is some internal degree of freedom at site $n$.
   In Boltzmann-style  arguments it is often assumed that each microstate is equally likely,  which corresponds to the probability $P_f^{N}$ on $A^N$ defined by  
  \beq
  P_f^{N}(\{\sg\})=|A|^{-N}=q^{-N},
  \eeq
   for each $\sg\in A^N$. This is the Bernoulli measure on $A^N$  induced by the \emph{flat prior} $p=f$ on $A$, 
   \beq
f(a)=1/q, \label{fA}
\eeq
 for each $a\in A$. More generally, $P^{N}_p$  is the Bernoulli measure on $A^N$  induced by some probability distribution $p$ on $A$; that is, the product measure of $N$ copies of $p$; some people write  $P^{N}_p=p^{\times N}$. 
This extends to the idealized case $A^{\omega}$, as follows. For $\sg\in A^N$ we define 
\beq
[\sg]:=\sg A^{\om}=\{s\in A^{\om}\mid \sg\prec s\},
\eeq
where $\sg\prec s$ means that $s=\sg\ta$ for some $\ta\in A^{\om}$ (in words: 
 $\sg\in A^*$ is a \emph{prefix} of $s\in A^{\om}$).
 On these basic measurable (and open) sets we define a probability measure
$P^{\omega}_p$ by 
$P^{\omega}_p([\sg])=P_p^N(\sg)$. In particular, $P^{\omega}_f([\sg])=|A|^{-N}$. 
It  is important to keep track of $p$  \emph{even if it is flat}: making no (apparent) assumption (which $p=f$ is often taken to be) is an important assumption!

 For example, Boltzmann's (1877) famous counting argument really reads as follows (Ellis, 1995; Dembo \& Zeitouni, 1998; Austin, 2017;  Dorlas, 2022).
 The formula 
 \beq
 S = k \log W\eeq
  on Boltzmann's grave should more precisely be something like
 \begin{equation}
S_B^{N}(\mu) = \log W^{N}(\mu), \label{grave}
\end{equation}
where I omit the constant $k$ and take $\mu\in\mathrm{Prob}(A)$ to be the relevant argument of the (extensive) Boltzmann entropy $S_B^{N}$ (see below). Furthermore, $W^{N}(\mu)$ is the probability (``Wahrscheinlichkeit'') of $\mu$, which Boltzmann, assuming the flat prior \er{fA} on $A$, took as
\begin{equation}
 W^{N}(\mu)=\frac{N(\mu)}{|A|^N}, \label{WNmu}
\end{equation}
where $N(\mu)$ is the number of microstates $\sg\in A^N$ whose corresponding  \emph{empirical measure}
\begin{equation}
L_N(\sg)=\frac{1}{N}\sum_{n=0}^{N-1} \dl_{\sg(n)}, \label{LN}
\end{equation}
equals $\mu$. Here, for any $b\in A$,  $\dl_b\in \mathrm{Prob}(A)$ is the point measure at $b$, i.e.\ $\dl_b(a)=\dl_{ab}$, the Kronecker delta. The number $N(\mu)$  is only nonzero if $\mu\in \mathrm{Prob}_N(A)$,
which consists of all probability distributions $\nu$ on $A$ that arise as $\nu=L_N(\sg)$ for some $\sg\in A^N$. This, in turn, means that $\nu(a)=\nu'(a)/N$ for some $\nu'(a)\in\{0, \ldots, N\}$, with $\sum_{a\in A} \nu'(a)=N$. In that case,
\begin{equation}
N(\mu)=\frac{N!}{\prod_{a\in A} (\mu'(a)!)}.
\end{equation}
The term $|A|^{-N}$ in \er{WNmu} of course equals $P^{N}(\sg)$ for any $\sg\in A^N$ and hence certainly for any $\sg$ for which $L_N(\sg)=\mu$. For such $\sg$, for general Bernoulli measures $P_p$ on $A^N$ we have
\begin{equation}
P^{N}_p(\sg)=e^{N\sum_{a\in A} \mu(a)\log p(a)}=e^{-N(h(\mu)+I(\mu|p))},
\end{equation}
 in terms of the \emph{Shannon entropy} and the \emph{Kullback--Leibler distance} (or \emph{divergence}),
 given by
\begin{align}
h(\mu)&:=-\sum_{a\in A} \mu(a)\log\mu(a); \label{SE}\\
I(\mu|p)&:=\sum_{a\in A} \mu(a)\log\left(\frac{\mu(a)}{p(a)}\right),
\label{KL}
\end{align}
respectively. These are simply related: for the flat prior \er{fA} we have 
\begin{align} I(\mu|f)=-h(\mu)+\log |A|. \label{Imuf}
\end{align}
In general, computing $S_B^{N}(\mu)$ from \er{grave} and, again assuming $L_N(\sg)=\mu$, we obtain
\beq
W^{N}(\mu)=P_p^{N}(L_N=\mu)=N(\mu)P_p^{N}(\sg)=\frac{N! \, e^{-N(h(\mu)+I(\mu|p))}}{\prod_{a\in A} (\mu'(a)!)},
\eeq
from which Stirling's formula (or the technique in Dembo \& Zeitiouni, 1998, \S 2.1) gives
\begin{equation}
s_B(\mu|p):=\lim_{N\raw\infty}\frac{S_B^{N}(\mu_N)}{N}=-I(\mu|p), \label{BE}
\end{equation}
where $\mu_N\in \mathrm{Prob}_N(A)$ is any sequence of probability distributions on $A$ that (weakly) converges to $\mu\in \mathrm{Prob}(A)$, i.e.\ the variable in $s_B(\cdot|p)$. For the flat prior  \er{fA}, eq.\ \er{Imuf} yields
\beq
s_B(\mu|f)=h(\mu)-\log |A|. \label{SBflat}
\eeq
As an aside, note that the Kullback--Leibler distance or relative entropy  \er{KL} is defined more generally for probability measures $\mu$ and $p$
 on some measure space $(A,\Sigma)$. As usual, we write
  $\mu \ll p$ iff $\mu$ is absolutely continuous with respect to $p$, i.e.\  $p(B)=0$ implies $\mu(B)=0$ for $B\in\Sg$. In that case,  the Radon--Nikodym derivative
 $d\mu/dp$ exists, and one has 
 \beq
 I(\mu|p):=\int_A dp\, \frac{d\mu}{dp}\log\left(\frac{d\mu}{dp}\right).
 \eeq
 If  $\mu$ is not absolutely continuous with respect to $p$, one puts $I(\mu|p):=\infty$. 
 The nature of  the empirical measure \er{LN} and the Kullback--Leibler distance \er{KL}
 comes out well in hypothesis testing. In order to test the hypothesis $H_0$ that $\mu=\mu_0$ by an $N$-fold trial $\sg\in A^N$, one  accepts  $H_0$ iff $I(L_N(\sg)|\mu_0)<\eta$, for some $\eta>0$. This test is optimal in the sense of Hoeffding, see Dembo \& Zetoumi (1998), \S 3.5. We now return to the main story.

The stochastic process $X_N:\Om_N\raw\mathcal{X}$ whose large fluctuations are described by  \er{BE} is 
\begin{align}
\mathcal{X}=\mathrm{Prob}(A); && \Om_N=A^N; && P_N=P^N_p; && X_N= L_N.
\end{align}
Then $L_N\raw p$ almost surely, and large fluctuations around this  value are described by 
\begin{align}
\lim_{N\raw\infty}\frac{1}{N}\log P_p^{N}(L_N\in \Gm)=-I(\Gm|p):= -\inf_{\mu\in\Gm}I(\mu|p)=\sup_{\mu\in\Gm} s_B(\mu|p), \label{LD1}
\end{align}
where $\Gm\subset \mathrm{Prob}(A)$ is open, or more generally, is such that $\Gm\subseteq\ovl{\mathrm{int(\Gm)}}$. Less precisely, 
\begin{align}
P_p^{N}(L_N\in \Gm)\approx e^{-NI(\Gm|p)}&& \mathrm{as}\:\: N\raw\infty,
\end{align} which implies that $P_p^{N}(L_N\in \Gm)\approx 1$ if $p\in\Gm$, whereas  $P_p^{N}(L_N\in \Gm)$ is exponentially damped if $p\notin\Gm$.
 Note that the rate function $\mu\mapsto I(\mu|p)$  defined in \er{KL} and \er{LD1} is  \emph{convex} and \emph{positive}, whereas the entropy  \er{BE} is \emph{concave} and \emph{negative}.
 Thus the former is to be \emph{minimized},  its \emph{infimum} (even minimum) over $\mu\in\mathrm{Prob}(A)$ being zero at $\mu=p$, whereas the latter is to be \emph{maximized}, its \emph{supremum} (even maximum) at the same value being zero.
The first term in \er{SBflat} hides the negativity of the Boltzmann entropy (here for a flat prior), but the second term drives it below zero. Positivity of $I(\mu|p)$ follows from (or actually \emph{is}) the Gibbs inequality.
Eq.\ \er{LD1}  is a special case of \emph{Sanov's theorem}, which works for arbitrary Polish spaces (instead of our finite set $A$); see Ellis (1985) or Dembo \& Zeitouni (1998). Einstein (1909) computes the probability of large fluctuations of the energy, rather than of the empirical measure, as in Boltzmann (1877), but these are closely related.

 Interpreting $A$ as a set of energy levels, the relevant stochastic process $X_N:\Om_N\raw\mathcal{X}$ still has 
\beq
(\Om_N, P_N)=(A^N, P^N_f),\eeq  but this time 
\beq
X_N=E_N=\frac{1}{N}\sum_{n=0}^{N-1} \sg(n),
\eeq taking values in $\mathcal{X}=\R$ (or some suitable finite subset thereof). This makes the relevant  entropy $s_C$ (which is the original  entropy from Clausius-style thermodynamics!) a function of $u\in\R$, interpreted as energy: instead of \er{LD1}, one obtains 
\begin{align}
\lim_{N\raw\infty}\frac{1}{N}\log P_p^{N}(E_N\in \Dl)&=\sup_{u\in\Dl} s_C(u|p); \\ s_C(u|p)&:= \sup_{\mu\in \mathrm{Prob}(A)}\left\{s_B(\mu|p)\mid \sum_{a\in A} \mu(a)\cdot a=u\right\},
\end{align}
which ``maximal entropy principle'' is a special case of \emph{Cram\'{e}r's theorem} (Ellis, 1985; Dembo \& Zeitouni, 1998). If  $\ovl{u}=\sum_{a\in A} p(a)\cdot a$ lies in $\Dl\subset\R$, then
$\log P_p^{N}(E_N\in \Dl)\approx 1$. If not, this probability is exponentially small in $N$. To obtain the classical thermodynamics of non-interacting particles (Dorlas, 2022), one may add that the free energy
\beq
f(\beta|p)=\log\left(\sum_{a\in A} p(a)e^{-\beta a}\right)\eeq
 is essentially the Fenchel transform (Borwein \& Zhu, 2005) of the entropy $s_C(u|p)$, in that
  \begin{align}
\beta  f(\beta|p)=\inf_{u\in\R}\{\beta u- s_C(u|p)\}; &&
  s_C(u|p)=\inf_{\beta\in\R}\{\beta u-\beta f(\beta|p)\}.
  \end{align} For $\beta>0$, the first equality is a refined version of ``$F=E-TS$''.
   \item In \emph{information theory} (Shannon, 1948; see also MacKay, 2003; Cover \& Thomas, 2006;  Lesne, 2014) the
``$N$''  in our diagram is the number of letters drawn from an alphabet $A$ by sampling a given probability distribution $p
\in\mathrm{Prob}(A)$, the space of all probability distributions on $A$. 
So each microstate $\sg\in A^N$ is a word with $N$ letters.
The \emph{entropy} of $p$, i.e.
 \begin{equation}
h_2(p):=-\sum_{a\in A} p(a)\log_2 p(a) =\sum_{a\in A} p(a)I_2(a),\label{SE2}
\end{equation}
plays a key role in Shannon's approach. It is the expectation value $h_2(p)=\la I\ra_p$ of the function
 \begin{equation}
I_2(a):=-\log_2 p(a),\label{SI}
\end{equation}
 interpreted as the \emph{information} contained in $a\in A$, relative to $p$. This interpretation is evident for the flat distribution $p=f$ on an alphabet with $|A|=2^n$ letters, in which case $I_2(a)=n$ for each $a\in A$, which is the minimal number of bits needed to (losslessly) encode $a$. 
 
  The general case is covered by the  \emph{noiseless coding theorem} for prefix (or uniquely decodable) codes.   A map $C:A\raw 2^*$ is a \emph{prefix code} if it is injective and $C(a)$ is never a prefix of $C(b)$ for any $a,b\in A$, that is,  there is no $\ta\in 2^*$ such that $C(b)=C(a)\ta$. A prefix code is uniquely decodable.
 Let $C:A\raw 2^*$ be a prefix code,
 let $\ell(C(a))$ be the length of the codeword $C(a)$, with expectation 
 \beq
 L(C,p)=\sum_{a\in A}p(a)\ell(C(a)).\eeq
 An \emph{optimal} code minimizes this. Then:
 \begin{enumerate}
\item Any prefix code satisfies $h_2(p)\leq L(C,p)$;
\item There exists an optimal prefix code $C$, which satisfies $L(C,p)\leq h_2(p)+1$.
\item  One has  $h_2(p)=  L(C,p)$ iff $\ell(C(a))=I_2(a)$ for each $a\in A$ (if this is possible).
\end{enumerate}
 Of course, the equality $\ell(C(a))=I_2(a)$ can only be satisfied if $p(a)=2^{-k}$ for some integer $k\in\N$. Otherwise, one can find a code for which $\ell(C(a))=[I_2(a)]$, the smallest integer $\geq I_2(a)$. See e.g.\ Cover \& Thomas (2006), \S 5.4.

Thus the information content $I_2(a)$ is approximately the length of the code-word $C(a)$ in some optimal coding $C$. Passing to our case of interest of $N$-letter words over $A$, in case of a memoryless source one simply has the Bernoulli measure $P_p^{N}$ on $A^N$, with entropy
 \begin{equation}
H_2(P_p^{N})=-\sum_{\sg\in A^N} P_p^{N}(\sg)\log_2 P_p^{N}(\sg)=Nh_2(p).\label{NH2}
\end{equation}
Extending the letter-code $C:A\raw 2^*$ to a word-code $C^N:A^N\raw 2^*$ by concatenation, i.e.\ $C^N(a_{i_0} \cdots a_{i_{N-1}}) =
C(a_{i_0}) \cdots C(a_{i_{N-1}})$, and replacing $L(C^N, P_p^{N})$, which diverges as $N\raw\infty$, by the average \emph{codeword length per symbol} $L(C^N, P_p^{N})/N$, an optimal code $C$ satisfies
\begin{equation}
\lim_{N\raw\infty} \frac{L(C^N, P_p^{N})}{N}=h_2(p).
\end{equation}
In what follows, the \emph{Asymptotic Equipartition Property} or AEP
will be important. In its (probabilistically) weak form, which is typically used in information theory, this  states that 
\begin{align}
\forall_{\varep>0} \lim_{N\raw\infty} P_p^{N}\left(\left\{\sg\in A^N\mid  P_p^{N}(\sg)\in [2^{-N(h_2(p)+\varep)}, 2^{-N(h_2(p)-\varep)}]
  \right\}\right)=1.
\end{align}
Its strong form, which is the (original) \emph{Shannon--McMillan--Breiman theorem},  reads
\begin{align}
P_p^{\om}\left(\left\{s\in A^{\om}\mid \lim_{N\raw\infty} -\frac{1}{N}\log_2  P_p^{N}(s_{|N})=h_2(p)  \right\}\right)=1. \label{SMB}
\end{align}
Either way,  the idea is that for large $N$, w.r.t.\  $P_p^{N}$ ``most'' strings $\sg\in A^N$ have ``almost'' the same probability
$2^{-N h_2(p)}$, whilst the others are negligible  (Austin, 2017, lecture 2). For $A=2$ with flat prior $p=f$ this yields a tautology: \emph{all} strings $\sg\in 2^N$ have $P_f^{N}(\sg)=2^{-N}$. See e.g.\ Cover \& Thomas (2006), \S 3.1 and \S16.8. The strong form follows from ergodic theory, cf.\ \er{SMB2}.
  \item In \emph{dynamical systems} along the lines of the ubiquitous Kolmogorov (1958),
 one starts with a triple $(X,P,T)$,
  where $X$--more precisely $(X,\Sg)$, but I usually suppress the $\sg$-algebra $\Sg$--is a measure space, $P$ is a probability measure on $X$ (more precisely, on $\Sg$), and $T:X\raw X$ is a measurable (but not necessarily invertible) map,   required to preserve $P$ in the sense that $P(T\inv B)=P(B)$ for any $B\in\Sg$. A measurable coarse-graining \er{Xdu} defines a map
\begin{align} \xi: X\raw A^{\om}; &&  \xi(x)_n=a\in A \:\:\mbox{ iff } \:\:T^nx\in X_a, \label{defxi}
\end{align}
in terms of which the given triple  $(X,P,T)$ is coarse-grained by a new triple $(A^{\om},  \xi_*P, S)$.
Here  $\xi_*P(B')=P(\xi\inv B)$ is the induced probability on $A^{\om}$, whilst  $S$ is  the (unilateral) \emph{shift}
\begin{align}
S: A^{\om}\raw A^{\om}; && (Ss)_n:=s_{n+1} \:\:\: (n=0,1,\ldots)
\end{align}
A fine-grained path $(x, Tx, T^2x, \ldots)\in X^{\om}$ is coarse-grained to $\xi(x)\in A^{\om}$, and truncating the latter at $t=N-1$ gives $\xi(x)_{|N}\in A^N$. Hence the configuration in the picture  states that
our particle starts from $x\in X_{a(0)}\subset X$ at $t=0$, moves to $Tx\in X_{a(1)}\subset X$ at $t=1$, etc.,
 and  at time $t=N-1$ finds itself at $T^{N-1}x\in  X_{a(N-1)}\subset X$. In other words, a coarse-grained path $\sg\in A^N$ tells us exactly  that $T^nx\in X_{\sg_n}$, for $n=0,1,\ldots, N-1$ ($\sg_n\in A$). Note that the shift satisfies
\beq
S\circ \xi=\xi\circ T,\eeq
so if $\xi$ were invertible, then nothing would be lost in coarse-graining; using the bilateral shift on $2^{\Z}$ instead of the unilateral one in the main text, this is the case for example with the Baker's map on $X=[0,1)\x [0,1)$  with $A=2$ and partition $(X_1=\{[0,1/2]\x[0,1)$, $X_2=(1/2,1)\x [0,1)\})$.
 The point of Kolmogorov's approach is  to refine the partition \er{Xdu}, which I now denote by 
 \beq
 \pi=\{X_a, a\in A\}\subset \Sg\subset P(X),\eeq
   to a finer partition $\pi^N=\{X_{\sg}, \sg\in A^N\}\subset \Sg$ of $X$, which consists of all non-empty subsets 
 \beq
 X_{\sg_0\cdots\sg_{N-1}}:=X_{\sg_0}\cap T\inv X_{\sg_1}\cap \cdots \cap T^{-(N-1)}X_{\sg_{N-1}} .
 \eeq
 Indeed, if we know $x$, then we know both the (truncated) fine- and coarse-grained paths 
 \begin{align}
 (x, Tx, \ldots, T^{N-1}x)\in X^N;&& \xi(x)_{|N}=\sg_0\cdots\sg_{N-1}\in A^N && (T^nx\in X_{\sg_n}, n\in N).
 \end{align}
 But if we just know that $x\in X_a$, we cannot construct even the coarse-grained path $\xi(x)_{|N}$. To do so, we must know that $x\in  X_{\sg}$, for some $\sg= \sg_0\cdots\sg_{N-1}\in A^N$ (provided $X_{\sg}\neq\emptyset$). 
 
 In other words,  the unique element $\pi^N(x)=X_{\sg(x)}$ of the partition $\pi^N$ that contains $x$, bijectively corresponds to a coarse-grained path $\sg(x)\in A^N$, and hence we may take
$P(\pi^N(x))$ to be the  probability of the coarse-grained path $\sg(x)$. This suggests an 
information function 
\beq
  I_{(X,P,T,\pi^N)}(x):=-\log_2P (\pi^N(x)),
\eeq cf.\ \er{SI}, 
and, as in \er{SE2},  an average (= expected) information or entropy function
\beq
H_{(X,P,T,\pi^N)}:= \la I_{(X,P,T,\pi^N)}\ra_P=\int_X dP(x)\, I_{(X,P,T,\pi^N)}(x)=
-\sum_{Y\in\pi^N} P(Y) \log_2(P(Y)). \label{ppKS}
\eeq
As $H_{(X,P,T,\pi^{M+N})}\leq H_{(X,P,T,\pi^M)}+ H_{(X,P,T,\pi^N)}$,
this (extensive) entropy has an (intensive) limit 
\beq
h_{(X,P,T,\pi)}:= \lim_{N\raw\infty} \frac{1}{N} H_{(X,P,T,\pi^N)}, \label{preKS}
\eeq
in turns of which the \emph{Kolmogorov--Sinai entropy} of our system $(X,P,T)$ is defined by
\begin{equation}
h_{(X,P,T)}:= \sup_{\pi} h_{(X,P,T,\pi)},\label{defKSE}
\end{equation}
where the supremum is taken over all finite measurable partitions of $X$, as above. 

 We say that  $(X,P,T)$ is \emph{ergodic} if for every $T$-invariant set $A\in\Sg$ (i.e.\ $T\inv A=A$), either $P(A)=0$ or $P(A)=1$. For later use I now state a number of equivalent conditions, each of which holds  iff $(X,P,T)$ is ergodic (Viana \& Oliveira, 2016, \S 4.1).
 For $P$-almost every $x$:\begin{align}
 \lim_{N\raw\infty}\frac{1}{N} \sum_{n=0}^{N-1}\dl_{T^nx}&=P && (\mbox{weakly in } \mathrm{Prob}(X)\label{Erg1};\\
 \lim_{N\raw\infty}\frac{1}{N} \sum_{n=0}^{N-1}f(T^nx)&=\int_X dP\,  f
 &&  \mbox{for each } f\in L^1(X,P); \label{Erg2}\\
\lim_{N\raw\infty}\frac{1}{N} |\{n\in\{0, \ldots, N-1\}: T^nx\in B\}|&=P(B);
 &&  \mbox{for each } B\in\Sg. \label{Erg3}
\end{align}
The empirical measure \er{LN} is a special case of \er{Erg1}. Eq.\ \er{Erg2} is 
a special case of Birkhoff's  ergodic theorem; eq.\  \er{Erg2} is Birkhoff's theorem \emph{assuming ergodicity}. In general, the l.h.s.\ is in $L^1(X,P)$ and is not constant $P$-a.e. 
Each of these is a corollary of the others, e.g.\ \er{Erg1} and \er{Erg2} are basically the same statement, and one obtains \er{Erg3} from \er{Erg2} by taking $f=1_B$. Note that 
the \emph{apparent} logical form of \er{Erg2} is: `for all $f$ and all $x$', which suggests that the universal quantifiers can be interchanged, but this is false: the \emph{actual} logical for is: `for all $f$ there exists a set of $P$-measure zero', which in general cannot be interchanged (indeed, in standard proofs the measure-zero set explicitly depends on $f$). Nonetheless, in some cases a measure zero set independent of $f$ can be found, e.g.\ for compact metric spaces and continuous $f$, cf.\   Viana \& Oliveira (2016), Theorem 3.2.6. 
Similar $f$-independence  will be true for the computable case reviewed below, which speaks in their favour. Likewise for \er{Erg3}.

Eq.\ \er{Erg2} implies the general Shannon--McMillan--Breiman theorem, which implies the previous one \er{SMB} for information theory by taking $(X=A^{\om}, P=P_p^{\om}, T=S)$. Namely:\begin{Theorem}\label{myT1}
If $(X,P,T)$ is ergodic, then for $P$-almost every $x\in X$ one has
\beq
h_{(X,P,T,\pi)}= - \lim_{N\raw\infty} \frac{1}{N}  \log_2 P(\pi^N(x)). \label{SMB2}
\eeq
\end{Theorem}
See e.g.\ Viana \& Oliveira (2016), Theorem 9.3.1, proved in \S 9.3.1.
Comparing this with \er{preKS} and \er{ppKS}, the average value of the information 
$I_{(X,P,T,\pi^N)}(x)$ w.r.t.\  $P$ can be computed from its value at a single point $x$, as long as this point is ``typical''. As in the explanation of the original  theorem in information theory, eq.\ \er{SMB2} implies that all typical paths (with respect to $P$) have about the same probability $\approx \exp(-Nh_{(X,P,T,\pi)})$.

  See  Sinai (1989) and more generally  Charpentier, Lesne, \& Nikolski (2007) for Kolmogorov's contributions to dynamical systems and ergodic theory. Relevant textbooks include for example Collet \& Eckmann (2006),  Castiglione \emph{et al.} (2008), and Viana \& Oliveira (2016).
  \end{itemize}
 \section{$P$-Randomness}\label{AR}
 The concept of $P$-randomness (where $P$ is a probability measure on some measure space $(X,\Sg)$) was introduced by \ML\ (1966) for the  case $X=2^{\om}$ and $P=P_f^{\om}$, i.e.\ the unbiased Bernoulli measure on the space of infinite coin flips. Following Hertling \& Weihrauch (2003), \S 3, a more general definition  of $P$-randomness which is elegant and  appropriate to my goals is as follows:
\begin{Definition}\label{defCR}
\begin{enumerate}\item[]
\item A topological space $X$ is \emph{effective} if it has a countable base $\mathcal{B}\subset\CO(X)$ with a bijection 
\beq
B:\N\stackrel{\cong}{\raw} \mathcal{B}.
\eeq An \emph{effective probability space}
$(X,B,P)$ is an effective topological space $X$  with a Borel probability measure $P$, i.e.\ defined on the open sets $\CO(X)$.
\item An open set $V\in\CO(X)$ as in 1.\ is \emph{computable} if  for some computable function $f:\N\raw\N$,
\beq
V=\bigcup_{n\in\N} B(f(n)).
\eeq 
Here $f$ may be  assumed to be total without loss of generality. In other words,
 \beq
 V=\bigcup_{n\in E} B(n)
 \eeq
  for some c.e.\ set $E\subset\N$ (where c.e.\ means \emph{computably enumerable}, i.e.\ $E\subset \N$ is the image of a total computable function $f:\N\raw\N$). 
\item A \emph{sequence} $(V_n)$ of opens $V_n\in\CO(X)$ is \emph{computable} if
\beq
V_n=\bigcup_{m\in\N} B(g(n,m))
\eeq for some  (total) computable function $g:\N\x\N\raw\N$; that is,
\beq
V_n=\bigcup_{m\mid (n,m)\in G} B(m)
\eeq
 for some c.e.\ $G\subset\N^2$. Without loss of generality we may and will assume that the (double) sequence $V_{(n,m)}=B(n)\cap B(m)$ is computable.
\item A  (randomness) \emph{test} is a computable sequence  $(V_n)$ as in 3.\ for which  for all $n\in\N$ one has
\beq
P(V_n)\leq 2^{-n}.\label{MLbound}
\eeq
One may (and will) also assume without loss of generality that   for all $n$ we have
 \beq
 V_{n+1}\subset V_n.\eeq
\item A point $x\in X$ is \emph{$P$-random} if $x\notin N$ for any subset $N\subset X$ of the form \beq
N=\bigcap_n V_n,
\eeq
 where $(V_n)$ is some test (since $P(\bigcap_n V_n)=0$, such an $N$ is called an 
\emph{effective null set}).
\item A   measure $P$ in an effective probability space $(X,B,P)$  is \emph{upper semi-computable} if the set
\begin{equation} U(P):= 
\left\{(F,q)\in P_f(\N)\x\Q\mid P\left(\bigcup_{n\in F}B(n)\right)<q\right\} \label{defUP}
\end{equation}
is c.e.\ (relative to some computable isomorphisms $P_f(\N)\cong\N$ and $\Q\cong\N$). Also, $P$ is  \emph{lower semi-computable} if the set $L(P)$, defined like \er{defUP} with $>q$ instead of $<q$, 
is c.e. Finally, $P$ is \emph{computable} if it is  upper and lower semi-computable,
 in which case $(X,B,P)$ is called a \emph{computable probability space (and similarly for upper and lower computability)}. 
  \end{enumerate}
\end{Definition} \noindent
Note that parts 1 to 5 do not impose any computability requirement on $P$, but even so it easily follows that $P(R)=1$, where $R\subset X$ is the set of all $P$-random points in $X$. 
However, if $P$ is upper semi-computable, one has a generalization of a further central result of \ML\ (1966), p.\ 605.
\begin{Definition}
A \emph{universal test} is a test $(U_n)$ such that for any  test $(V_n)$ there is a constant $c=c(U,V)\in\N$ such that for each $n\in\N$ we have $V_{n+c}\subset U_n$. 
\end{Definition}\noindent
Universal tests, then, exist provided  $P$ is upper semi-computable, which in turn implies that $x\in X$ is $P$-random iff $x\notin U$. See Hertling \& Weihrauch (2003), Theorem 3.10. 
Compared with the computable metric spaces of Hoyrup \& Rojas (2009), which for all purposes of this paper could have been used, too, Hertling \& Weihrauch (2003), whom I follow here,  avoid the choice of a countable dense subset of $X$. The latter is unnatural already in the case $X=A^{\om}$, where $\sg\in A^*\cong\N$ has to be injected into $A^{\om}$ via a map like $\sg\mapsto \sg a^{\om}$ for some fixed $a\in A$ (where $a^{\om}$ repeats $a$ infinitely often). On the other hand, the map $A^*\raw \CO(A^{\om})$, $\sg\mapsto [\sg]$, where $[\sg]=\sg A^{\om}$, is quite natural (here $\CO(X)$ is the topology of $X$). 
If $P$ is computable as defined in clause 6,  $P$ is a computable point in the effective space of all probability measures on $X$ (Hoyrup \& Rute, 2021), where a point $x$ in an effective topological space is deemed \emph{computable} if $\{x\}=\cap_n V_n$ for some computable sequence $V_n$.

A key example is $X=A^{\om}$ over a finite alphabet $A$, with topology $\CO(X)$ generated by the cylinder sets $[\sg]=\sg A^{\om}$, where $\sg\in A^N$ and $N\in\N$. 
The usual lexicographical order on $A^*$ then gives a bijection $L:\N\stackrel{\cong}{\raw} A^*$, and hence 
a numbering 
\begin{align}
B:\N\stackrel{\cong}{\raw} \mathcal{B}; && n\mapsto [L(n)]= L(n)A^{\om}.\label{BL}
\end{align}
The Bernoulli measures $P^{\om}_p$ on $A^{\om}$ then have the same computability properties as $p\in \mathrm{Prob}(A)$. In particular, the flat prior $f$ makes $P^{\om}_f$ computable, and in case that $A=2$, the computability properties of $p\in[0,1]\cong \mathrm{Prob}(2)$ are transferred to $P^{\om}_p$. This is all we need for my  main theme.

In case of a  flat prior $f$ on $A$, the above notion of randomness of sequences in $A^{\om}$ is equivalent to the  definition in \ML\ (1966), which I will now review, following Calude (2002). Though equivalent to the definition of \ML\ random sequences in books like 
   Li \& Vit\'{a}nyi (2008), Nies (2009), and  Downey \& Hirschfeldt (2010), the construction in Calude (2002) is actually closer in spirit to \ML\ (1966) and has the advantage of being compatible with Earman's principle (see below) even before redefining randomness in terms of Kolmogorov complexity. The definition in the other three books just cited lacks this feature. Calude's definition is based on first defining random \emph{strings} $\sg\in A^*$. Since $A^*$ has the discrete topology, we simply take $\CB$ to consist of all singletons $\{\sg\}$, $\sg\in A^*$, with $B=L$ as in  \er{BL}. Since unlike $A^N$ or $A^{\om}$ the set $A^*$ does not carry a useful probability measure $P$, we replace \er{MLbound} by
   \begin{equation}
P_f^N(V_n\cap A^N)\leq  \frac{ |A|^{-n}}{|A|-1}.
\label{MLR1}
\end{equation}   \vspace{-5mm}
\begin{Definition}\label{defST}
A \emph{sequential test} is a computable sequence  $(V_n)$ of subsets $V_n\subset A^*$ such that:
\begin{enumerate}
\item  The inequality \er{MLR1} holds;
\item  $V_{n+1}\subset V_n$ (as in Definition \ref{defCR}.4);
\item $\sg\in V_n$ and $\sg\prec\ta$ imply $\ta\in V_n$ (i.e.\ extensions of $\sg\in V_n$ also belong to $V_n$). 
\end{enumerate}
 \end{Definition}
 Since \er{MLR1} is the same as
   \begin{equation}
|V_n\cap A^N| \leq  \frac{ |A|^{N-n}}{|A|-1},\label{Cal2}
\end{equation}   
Via eq.\er{Cal2}, Definition \er{defST}  implies that $V_n\cap A^N=\emptyset$ for all $N<n$.
A simple example of a sequential test for $A=2$ is $V_n=\{\sg\in 2^*\mid 1^n\prec\sg\}$, i.e.\ the set of all strings starting with $n$ copies of 1. There exists a \emph{universal} sequential test $(U_n)$ such that for any sequential test $(V_n)$ there is a $c=c(U,V)\in\N$ such that for each $n\in\N$ we have $V_{n+c}\subset U_n$. See Calude (2002), Theorem 6.16 and Definition 6.17. 
For this (or indeed any) test $U$ we  define $m_U(\sg):=0$ if $\sg\notin U_1$, and otherwise
\begin{align}
m_U(\sg):=\max\{n\in\N\mid \sg\in U_n\}.
\end{align}
By the comment after \er{Cal2} we have $m_U(\sg)\leq|\sg|<\infty$, since $\sg\in A^N$ for some $N$. 
If $m_U(\sg)<q$ for some $q\in\N$, then $\sg\notin U_q$ by definition.   Since $U_{n+1}\subset U_n$,
 this implies $\sg\notin U_{q'}$ for all $q'>q$, so that also $m_U(\sg)<q'$. But as we have just seen, we may restrict these values to $q'\leq|\sg|$.
 \begin{Definition}\label{Caldef} 
\begin{enumerate}
\item A string $\sg\in A^*$ is \emph{$q$-random} (for some $q\in\N$) if 
$m_U(\sg)<q\leq |\sg|$.
\item A  sequence $s\in A^{\om}$ is \emph{Calude random} (with respect to $P=P_f^{\om}$) if there is a constant $q\in\N$ such that each finite segment $s_{|N}\in A^N\subset A^*$ is $q$-random, i.e., such that for all $N$,
\beq
m_U(s_{|N})<q. \label{EP1}
\eeq
\end{enumerate}
 \end{Definition}\noindent
Note that  the \emph{lower} $q$ is, the \emph{higher} the randomness of $\sg$, as it lies in \emph{fewer} sets $U_n$. 
It is easy to show that Calude randomness is equivalent to any of
the following three conditions (the third of these is taken as the definition of randomness by Calude (2002), Definition 6.25):
\begin{align}
\lim_{N\raw\infty} m_U(s_{|N})<\infty; &&\lim\sup_{N\raw\infty} m_U(s_{|N})<\infty;\\
\lim_{N\raw\infty} m_V(s_{|N})<\infty && \mbox{for all sequential tests } (V_n).
 \label{MLR3}
\end{align}\vspace{-5mm}
\begin{Theorem}\label{T3.4}
A string $s\in A^{\om}$ is $P_f^{\om}$-random (cf.\ Definition \ref{defCR}.5) iff it is Calude random.
\end{Theorem}\noindent
This follows from Theorem 6.35 in  Calude (2002), \S 6.3, and Theorem \ref{Calthm} below. The point is that
randomness of sequences $s\in A^{\om}$ can  be expressed in terms of randomness of finite initial segments of $s$. This is also true via another (much better known) reformulation of \ML-randomness.
\begin{Definition}\label{Koldef} 
A  sequence $s\in A^{\om}$ is  \emph{Chaitin--Levin--Schnorr random} if there is a constant $c\in\N$
such that each finite segment $s_{|N}\in A^N$ is \emph{prefix Kolmogorov $c$-random}, in the sense that
  for all $N$,
\begin{equation}
K(s_{|N})\geq N-c.  \label{EP2}
\end{equation}
\end{Definition}\noindent
Here $K(\sg)$ is  the prefix Kolmogorov complexity of $\sg$ with respect to a fixed universal prefix Turing machine; changing this machine only changes the constant $c$ in the same way for all strings $\sg$ (which  makes the  value of $c$ somewhat arbitrary). Recall that  $K(\sg)$ of $\sg\in A^*$ is defined as the length of the shortest  program that outputs $\sg$ and then halts, running on a universal prefix Turing machine $T$ (i.e., the domain $D(T)$ of $T$
consists of a prefix subset of $2^*$, so if $x\in D(T)$ then $y\notin D(T)$ whenever $x\prec y$). Fix some  universal prefix Turing machine $T$, and
 define $K(\sg):=\min\{|x|: x\in 2^*, T(x)=\sg\}$.  Then $\sg\in A^*$ is \emph{$c$-prefix Kolmogorov random}, for some $\sg$-independent  $c\in\N$, if $K(\sg)\geq |\sg|-c$.  Note that Calude (2002) writes $H(\sg)$ for what I call $K(\sg)$, following
 Li \& Vit\'{a}nyi  (2008) and others. 
A key result  algorithmic randomness, then, is (e.g.\ Downey \& Hirschfeldt, Theorem 6.2.3):
\begin{Theorem}\label{Calthm}
A sequence $s\in A^{\om}$ is $P_f^{\om}$-random iff it is Chaitin--Levin--Schnorr random.
\end{Theorem}
\noindent   According to Kjos-Hanssen \& Szabados (2011), p.\ 3308, footnote 1 ((references adapted):
\begin{quote}
\begin{small}
[Theorem 3]  was announced by Chaitin (1975) and attributed to Schnorr (who was the referee of the paper) without proof. The first published proof (in a form generalized to arbitrary computable measures) appeared in the work of G\'{a}cs (1979).
\end{small}
\end{quote} 
 Since Levin (1973) also states  Theorem \ref{Calthm}, 
the names \emph{Chaitin--Levin--Schnorr} seem fair.\\
 Hence both Definitions \ref{Caldef} and \ref{Koldef} are  compatible \emph{on their own terms} with 
  \emph{Earman's Principle}:\smallskip
  
  \begin{quote}
\begin{small}
While idealizations are useful and, perhaps, even essential to progress in physics, a sound principle of interpretation would seem to be that no effect  can be counted as  a genuine physical effect if it disappears
when the idealizations are removed.  (Earman, 2004, p.\ 191)
\end{small}
\end{quote}

By Theorem \ref{T3.4}, Definition \ref{Caldef} is a special case of Definition \ref{defCR}.5 and hence it 
 depends on the initial probability $P^{\om}_f$ on $A^{\om}$. On the other hand, both \er{Cal2} and the equivalence between  Definitions \ref{Caldef}  and \ref{Calthm} suggest that $P^{\om}_f$-randomness 
 does not depend on $P^{\om}_f$! To assess this, let us look at a version of Theorem \ref{Calthm} for arbitrary computable measures $P$ on $2^{\om}$ (Levin, 1973; G\'{a}cs, 1979).
\begin{Theorem}\label{LG}
Let $P$ be a computable probability measure on $2^{\om}$. Then $s\in 2^{\om}$ is $P$-random iff  there is a constant $c\in\N$ such that
  for all $N$,
\begin{equation}
K(s_{|N})\geq -\log_2( P([s_{|N}]))-c.  \label{EPLG}
\end{equation}
\end{Theorem} 
If $P=P_f^{\om}$, then $P([s_{|N}])=2^{-N}$,  and so \er{EPLG}  reduces to \er{EP2}. Thus the absence of a $P$-dependence in Definition \ref{Koldef} is only apparent, since it \emph{implicitly} depends on the  assumption $p=f$.  

\noindent It seems, then, that
 Kolmogorov did not achieve his goal of defining randomness in a non-probabilistic  way!
Indeed, note also that the definition of $K(\sg)$ depends on the hidden assumption that the length function $\sg\mapsto |\sg|$ on $2^*$ assigns equal length to 0 and 1 (Chris Porter, email June 13, 2023).

Another interesting example is Brownian motion, which is  related to binary sequences via
the \emph{random walk} ( M\"{o}rters \& Peres, 2010; McKean, 2014). Brownian motion may be defined as a Gaussian stochastic process $(B_t)_{t\in[0,1]}$ in $\R$ with variance $t$ and  covariance 
 $\la B_s B_t\ra =\min(s,t)$.
We also assume that  $B_0=0$. An equivalent axiomatization states that for each $n$-tuple $(t_1, \ldots t_n)$ with $0\leq t_1\leq \cdots\leq t_n$ the increments $B_{t_n}-B_{t_{n-1}}$, \ldots, $B_{t_2}-B_{t_1}$ are independent, that for each $t$ one has  
\beq
P(B_{t+h}-B_t\in [a,b])= (2\pi t)^{-1/2}\int_a^bdx\, e^{-x^2/2h},
\eeq
 and that $t\mapsto B_t$ is continuous with probability one. If we add that  $B_0=0$, these axioms imply 
 \beq
 P(B_t\in [a,b])= (2\pi t)^{-1/2}\int_a^bdx\, e^{-x^2/2t}.
 \eeq 

 We switch from $2=\{0,1\}$ to $\ut=\{-1,1\}$. Take $C[0,1]\equiv C([0,1],\R)$, seen as a Banach space in the supremum norm $\| f\|_{\infty}=\sup\{|f(x)|, x\in [0,1]\}$ and hence as a metric space (i.e.\ $d(f,g)=\| f-g\|_{\infty}$) with ensuing Borel structure.
For each $N=1,2,\ldots$, define a map
\begin{align}
R_N: \ut^N\raw C[0,1]; && R_N(\usg)(0):=0; && R_N(\usg)\left(\frac{n}{N}\right):=\frac{1}{\sqrt{N}}\sum_{k=1}^n \usg_k\:\:\:\: (n=1, \ldots, N),\end{align}
and $R_N(\usg)$ is defined at all other points $t\neq n/N$ of $[0,1]$ via linear interpolation (i.e.\ by drawing straight lines between $R_N(\usg)(n-1)$ and $R_N(\usg)(n)$ for each $n=1, \ldots, N$; I omit the formula). Thus $R_N(\usg)$ is a random walk with $N$ steps in which each time jump $t=0,1,\ldots N$ is compressed from unit duration to  $1/N$ (so that the time span $[0,N]$ becomes $[0,1]$), and each spatial step size is compressed from $\pm 1$ to $\pm 1/\sqrt{N}$. Now
equip $\ut^N$ with the fair Bernoulli probability measure $P^N_f$. Then $R_N$ induces a probability measure $P^N_W$ on $C[0,1]$ in the usual way, i.e.\ 
\beq
P^N_W(A)=P^N_f(R_N\inv(A)),
\eeq
 for measurable 
$A\subset C[0,1]$. The point, then, is that there is a unique probability measure $P_W$ on $C[0,1]$, called  \emph{Wiener measure}, such that $P_W^N\raw P_W$ weakly as $N\raw\infty$.
The concept of weak convergence of probability measures on (complete separable) metric spaces $X$ used here is defined as follows: a sequence $(P_N)$ of 
probability measures on $X$ converges weakly to $P$ iff 
\beq
\lim_{N\raw\infty} \int_X dP_N\, f=\int_X dP\, f,
\eeq for each $f\in C_b(X)$. This is equivalent to $P_N(A)\raw P(A)$ for each measurable $A\subset X$ for which $\partial A=\emptyset$.
 See Billingsley (1968) for both the general theory and its application to 
Brownian motion, which may now be realized on $(C[0,1], P_W)$ as $B_t=\mathrm{ev}_t$, where the evaluation maps are defined by
 \begin{align}
 \mathrm{ev}_t:C[0,1]\raw\R; && \mathrm{ev}_t(f)=f(t).
 \end{align}
  In fact, the set of all paths of the kind $R_N(\usg)$, $\usg\in\ut^N$, is uniformly dense
 in $C_0[0,1]$, the set of all $B\in C[0,1]$  that vanish at $t=0$ (on which $P_W$ is supported).
 Namely, 
 for $B\in  C_0[0,1]$ and  $N>0$, recursively define 
 \begin{align}
 t_1&:=\min\{t\in[0,1], |B(t)|=1/\sqrt{N}\}; \:\:\:\:
  t_2:=\min\{t\geq t_1, |B(t)-B(t_1)|=1/\sqrt{N}\}; \\
 t_{n}&:=\min\{t\geq t_{n-1}, |B(t)-B(t_{n-1})|=1/\sqrt{N}\},
 \end{align}
  until $n=N$. In terms of these, define $\usg^{(N)}\in\ut^N$ by $\usg^{(N)}_0=0$ and then, again recursively    until $n=N$,
   \begin{align}
 \usg^{(N)}_1=\sqrt{N} B(t_1), \usg^{(N)}_2=\sqrt{N} (B(t_2)-B(t_1)), && \ldots,  &&
 \usg^{(N)}_n=\sqrt{N} (B(t_{n})-B(t_{n-1})).
 \end{align}  
 Then $R_N(\usg^{(N)})\raw B$ as $N\raw\infty$, but $\|B-R_N(\usg^{(N)})\|_{\infty}$ is just $O(N^{-1/18})$, cf.\  McKean (2014), \S 6.4.1.
 
 This enables us to turn $(C[0,1], P_W)$ (with suppressed Borel structure given by the metric) into an effective probability space. In Definition \ref{defCR}.1, we take the  countable base $\CB$ to consist of all open balls with rational radii around points $R_N(\usg^{(N)})$, where $N\in \N_*$ and $\usg^{(N)}\in\ut^N$,
 numbered via lexicographical ordering of $\ut^*$ and computable ismorphisms $\Q^+\cong\N$ and $\N^2\cong\N$. 
 
 The following theorem (Asarin \& Prokovskii, 1986) characterizes the ensuing notion of $P_W$-randomness.
 See also  Fouch\'{e} (2000ab) and   Kjos-Hansen \& Szabados (2011). 
   \begin{Theorem}\label{AP}
 A path $B\in C[0,1]$ is $P_W$-random iff $B=\lim_{N\raw\infty} B_N$ (w.r.t.\ $\|\cdot\|_{\infty}$) \emph{effectively} for some sequence $B_N=R_N(\usg^{(N)})$ in $C[0,1]$ for which 
 there is a constant $c\in\N$ such that for all $N$, 
 \begin{equation}
K(\usg^{(N)})\geq N-c.  \label{EP3}
\end{equation} 
 \end{Theorem}
 \noindent Here \emph{effective} convergence $B_N\raw B$ means that 
 \beq
 \forall_{m\in\N_*}\exists_{N(m)}\forall_{N>N(m)} \| B-B_N\|_{\infty}<1/m,
 \eeq where $m\mapsto N(m)$ is computable (so $1/m\in\Q^+$ plays the role of  $\varep\in\R^+$).

 Compare with Definition \ref{Koldef}. It might be preferable if there were a single sequence $\underline{s}\in\ut^{\om}$ for which $K(s_{|N})\geq N-c$, cf.\ 
  \er{EP2}, but unfortunately this is not the case (Fouch\'{e}, 2000ab). Nonetheless, Theorem \ref{AP} is  satisfactory from the point of view of Earman's principle above, in that randomness of a Brownian path is characterized by randomness properties of its finite approximants $B_N$; indeed, each $\usg^{(N)}\in\ut^N$ is $c$-Kolmogorov random, even for the same value of $c$ for all $N$.   
 \section{From `for $P$-almost every $x$' to `for all $P$-random $x$'} \label{PP}
Although results of the kind reviewed here pervade the literature on algorithmic randomness (and, as  remarked in the Introduction, might be said to be a key goal of this theory), 
their importance for physics still remains to be explored. The idea is best illustrated by the following example, which was the first of its kind. For binary sequences in $2^{\om}$ equipped with the flat Bernoulli measure $P_f^{\om}\equiv P_{1/2}^{\om}$, see \er{fA} etc., the strong law of large numbers  (e.g.\ McKean, 2014, \S 2.1.2) states that
\begin{equation}
\lim_{N\raw\infty} \frac{1}{N}\sum_{n=0}^{N-1} s_n=1/2, \label{SLLN}
\end{equation}
\emph{for $P_f^{\om}$-almost every $s\in 2^{\om}$} (or: $P_f^{\om}$-almost surely). Recall that this means that there exists a measurable subset $A\subset 2^{\om}$ with $P_f^{\om}(A)=1$ such that \er{SLLN} holds for each $s\in A$ (equivalently:
there exists $B\subset 2^{\om}$ with $P_f^{\om}(B)=0$ such that \er{SLLN} holds for each $s\notin B$).
Theorems like this provide no information about $A$ (or $B$). 
 \ML\ randomness (cf.\   Definition \ref{defCR}) provides this  information (usually at the cost of additional computability assumptions), where I recall that,
 as explained more generally after Definition \ref{defCR},
 the set $R\subset X$ of all $P$-random elements in $X$  has $P(R)=1$.   In
the case at hand, the  computability assumption behind this result is  satisfied since we use a computable flat prior $p=f$ under which $(2^{\om}, P_f^{\om})$ is a computable probability space in the sense of Definition \ref{defCR}.
\begin{Theorem}
The strong law of large numbers \er{SLLN} holds for all 
 $P_f^{\om}$-random sequences $s\in 2^{\om}$.
 \end{Theorem}\noindent
  See  \ML\ (1966), p.\ 619, and in detail Calude (2002), Theorem 6.57.  The law of the iterated logarithm   (McKean, 2014, \S 2.3) also holds in this sense  (Vovk, 1987).  
  More generally, the classical theorem stating that $P_f^{\om}$-almost all sequences  $s\in A^{\om}$ are Borel normal can be sharpened to the statement that \emph{all $P^{\om}_f$-random sequences $s\in A^{\om}$ are Borel normal.} See Calude (2002), Theorem 6.61 (a sequence $s\in A^{\om}$ is called \emph{Borel normal} if each string $\sg\in A^N$ occurs in $s$ with the asymptotic relative frequency $|A|^{-N}$ given by $P_f^{\om}$). 
  The most spectacular result in this direction is arguably:
 \begin{Theorem}\label{MTT}
 Any $\sg\in A^*$ occurs infinitely often in every $P^{\om}_f$-random sequence $s\in A^{\om}$.
  \end{Theorem}
 \noindent  See  Calude (2002) Theorem 6.50; the original version of this
   ``Monkey typewriter theorem'' of course states  that
  any $\sg\in A^*$ occurs infinitely often in $P_f^{\om}$-almost all sequences  $s\in A^{\om}$. 
  But I wonder if this theorem matches Earman's principle: I see no interesting and valid version for finite strings.

 The proof of all such theorems, including those to be mentioned below, is by contradiction:
 $x$ \emph{not} having the property $\Phi(x)$ in question, e.g.\ \er{SLLN}, would make $x$ fail some randomness test. 
   
Interesting examples also come from analysis. The pertinent computable probability space is $([0,1],\lm)$,
where $\lm$ is Lebesgue measure,  and
for the basic opens $\CB$ in Definition \ref{defCR} one  takes open intervals with rational endpoints, suitably numbered (here I suppress the usual Borel $\sg$-algebra $B$ on $[0,1]$, which is generated by the standard topology). Alternatively,  the map
\begin{align}
2^{\om}\raw [0,1]; && s\mapsto \sum_{n=0}^{\infty} \frac{s_n}{2^{n+1}}, \label{notiso}
\end{align}
induces an isomorphism of probability spaces $(2^{\om},P_{1/2}^{\om})\stackrel{\cong}{\raw} ([0,1],\lm)$, though not a bijection of sets $2^{\om}\raw [0,1]$, since the dyadic numbers (i.e.\ $x=m/2^n$ for $n\in\N$ and $m=1,2, \ldots, 2^n-1$) have no unique binary expansions (the potential non-uniqueness of binary expansions is irrelevant for the purposes of this section, since dyadic numbers are not random).
Although \er{notiso} is not a homeomorphism, it nonetheless maps the usual $\sg$-algebra of measurable subsets of $2^{\om}$ to its counterpart for $[0,1]$.
By Corollary 5.2 in  Hertling \& Weihrauch (2003) we then have:
\begin{Theorem}\label{DT1}
Let $x=\sum_n s_n2^{-n-1}$. Then $x\in [0,1]$ is $\lm$-random iff  $s\in 2^{\om}$ is  $P_f^{\om}$-random.      \end{Theorem}
This matches Theorem \ref{AP} in reducing a seemingly different setting for randomness to the case of binary sequences. See  Hoyrup \& Rute (2021) for a general perspective on this phenomenon. 

One of the clearest theorems relating analysis to randomness in the spirit of our theme is the following  (Brattka, Miller, \& Nies, 2015, Theorem 6.7), which sharpens a classical result to the effect that
any function $f:[0,1]\raw\R$ of bounded variation is almost everywhere differentiable.
First, recall that $f$ has \emph{bounded variation} if there is a constant $C<\infty$ such that for any finite collection of points $0\leq x_0<x_1 \cdots< x_n<x_{n+1}\leq1$ one has $\sum_{k=0}^n| f(x_{k+1}-f(x_k)|<C$. By the Jordan decomposition theorem, this turns out to be the case iff  $f = g -h$ where $g$ and $h$ are non-decreasing. 
\begin{Theorem}\label{BVT}
If $f:[0,1]\raw\R$ is computable and has bounded variation, then $f$ is differentiable at any $\lm$-random $x\in[0,1]$. Moreover, $x\in[0,1]$ is $\lm$-random iff $f'(x)$ exists for every such $f$.
\end{Theorem}
Theorems like this  give us even more than we asked for (which was the mere ability to replace
 `for $P$-almost every $x$' by `for all $P$-random $x$'): they \emph{characterize} random points in terms of a certain property to be had by a specific class of computable functions. I here use the definition (or characterization) of computability due to Hertling \& Weihrauch (2003), Definition 4.2:
 if $(X,\CB)$ and $(X',\CB')$ are effective topological spaces (see Definition \ref{defCR}.1 above), then
 $f:X\raw X'$ is computable iff for each $U'\in\CB'$ the inverse image $f\inv(U')\subset X$ is open and computable (cf.\ Definition \ref{defCR}.3). 
 
 There is also a similar result in which bounded variation is replaced by absolute continuity. Theorem \ref{BVT} also has a counterpart in which $f$ is non-decreasing, but here the conclusion is that $f$
is \emph{computably} random instead of \ML-random (see Downey, Griffiths, \& Laforte (2004) for computable randomess, originally defined by Schnorr via martingales, which is  weaker than \ML\ randomness, i.e.\ 
\ML-randomness implies computable randomness).
Another classical result in the same direction returns \emph{Schnorr} randomness (recall that $x\in X$ is \emph{Schnorr random} if in Definition \ref{defCR}.4 we replace \er{MLbound} by  $P(V_n)=2^{-n}$; this gives fewer tests to pass, and hence, once again, a weaker sense of randomness than \ML\ randomness). 
\begin{Theorem}
If $f\in L^1[0,1]$ is computable, then 
$\lim_{h\raw 0} \frac{1}{2h} \int_{x-h}^{x+h} dy\, f(y)$ exists for all $\lm$-random $x\in [0,1]$, and
the above limit exists \emph{for each computable} $f\in L^1$ iff
 $x\in[0,1]$ is \emph{Schnorr} random. 
\end{Theorem}
\noindent See Brattka, Miller, \& Nies (2015), reviewed in  Hoyrup \& Rute (2021), \S 3.2.  
 
We now turn to ergodic theory. Here, my  favourite example  is the following.
 Recall the 
 equivalent characterizations of ergodicity stated in eqs.\ \er{Erg1} to \er{Erg3} in \S\ref{BEP}. 
\begin{Theorem}\label{CET}
Let $(X,P,T)$ be ergodic with $P$ and $T$ computable. Then \er{Erg1},  restricted to
\begin{equation}
 \lim_{N\raw\infty}\frac{1}{N} \sum_{n=0}^{N-1}\dl_{T^nx}(V)=P(V), \label{Erg1b} 
\end{equation}
for each computable open $V\subset X$ (cf.\ Definition \ref{defCR}.3),
 holds for all $P$-random $x\in X$. Moreover, $x\in X$ is $P$-random iff  \er{Erg1b} holds for every computable $T$ and every computable open $V\subset X$.
\end{Theorem}
\noindent See Galatolo, Hoyrup, \& Rojas (2010), Theorem 3.2.2, and Pathak, Rojas, \& Simpson (2014), Theorem 1.3.
 The first author to prove such results was V'yugin (1997). See also the reviews by Towsner (2020) and 
  V'yugin (2022). In Theorem \ref{CET} one could replace \er{Erg1b} with the property that $x$ satisfy (Poincar\'{e}) recurrence, in the sense that for each computable open $V\subset X$ (not necessarily containing $x$) there is some $n\in\N$ such that $T^n(x)\in V$. 
   If \er{Erg2} instead of \er{Erg1} is used, a  result like Theorem \ref{CET} obtains that characterizes Schnorr randomness. 
 The  Shannon--McMillan--Breiman theorem \er{SMB2} also falls under this scope.
 We say that a partition $\pi$ of $X$ is computable if each $X_A\subset X$ is a computable open set. The defining equation \er{Xdu} is then replaced by $P(\bigsqcup_{a\in A} X_{a})=1$.
  \begin{Theorem}\label{CMBb}
If $P$ and $T$ are computable and $T$ is ergodic, and also the partition $\pi$ of $X$ is computable, then for every $P$-random $x\in X$ one has
\beq
h_{(X,P,T,\pi)}= - \lim\sup_{N\raw\infty} \frac{1}{N}  \log_2 P(\pi^N(x)). \label{SMB2b}
\eeq
\end{Theorem}\noindent
See  Galatolo,  Hoyrup, \& Rojas (2010), Corollary 6.1.1 (note the lim sup here). 
Things become more interesting if we replace the  \emph{information function}  $-\log_2P (\pi^N(x))$ by the (prefix) \emph{Kolmogorov complexity} $K(\xi_N(x))$. Recall the map \er{defxi}, which we may truncate to maps
 \begin{align}
 \xi_N: X\raw A^N; && \xi_N(x)=\xi(x)_{|N},
 \end{align} so that $\xi_N(x)_n=a$ (for $n=0, \ldots, N-1$) identifies the subspace $X_a$ of the partition our particle occupies after $n$ time steps.
 Under the same assumptions as Theorem \ref{CMBb}, we then have:
 \beq
h_{(X,P,T,\pi)}=  \lim\sup_{N\raw\infty} \frac{1}{N} K(\xi_N(x)),\label{Brudno}
\eeq for
all $P$-random $x\in X$ (and hence for $P$-almost every $x\in X$), cf.\ \er{preKS}.
 See Brudno (1983), White (1993), Batterman \& White (1996), and  Galatolo,  Hoyrup, \& Rojas (2010).
  Taking the supremum over all computable partitions $\pi$,  \emph{the Kolmogorov--Sinai entropy of $(X,P,T)$ equals the limiting Kolmogorov complexity of any increasingly fine coarse-grained $P$-random path for $(X,T)$.}  
Note that the right-hand side of \er{Brudno} is independent of $P$, which the left-hand side is not; however, the condition for the validity of \er{Brudno}, namely that $x$ be $P$-random, depends on $P$. The equality
\begin{equation}
\lim\sup_{N\raw\infty} \frac{1}{N}\la K\circ \xi_N\ra_P=\lim\sup_{N\raw\infty} \frac{1}{N} K(\xi_N(x)),
\end{equation}
for all $P$-random $x\in X$  also illustrates our theme; it shows that (at least asymptotically) each $P$-random $x$ generates a course-grained path $\xi_N(x)$ that has  ``average'' Kolmogorov complexity. 

Applying these results to $X=A^{\om}$, with $T$ the unilateral shift and $P=P_p^{\om}$ the Bernoulli measure on $A^{\om}$ given by a probability distribution $p$ on some alphabet $A$, gives a similar expression for the Shannon entropy \er{SE}:
 for all $P_p$-random $s\in A^{\om}$ (and hence for $P_p$-almost every $s\in A^{\om}$),
\begin{equation}
 \lim_{N\raw\infty}\frac{1}{N} K(s_{|N})= h_2(p).  \label{folklore} 
\end{equation}
Note the $\lim$ instead of the $\lim\sup$, which is apparently justified in this special case.  Porter (2020) states my eq.\ \er{folklore} as his Theorem 3.2 and labels it ``folklore", referencing however Levin \& Zvonkin (1970) and Brudno (1978). See  also Schack (1998) and Gr\"{u}nwald \& Vit\'{a}nyi (2003) for further connections between entropy and algorithmic complexity. 

Finally, here are some nice examples involving Brownian motion. Three classical results  are:
  \begin{Theorem}\label{BM1}
  \begin{enumerate}
\item For $P_W$-almost every $B\in C[0,1]$ there exists $h_0>0$ such that 
\begin{equation}
|B(t+h)-B(t)|\leq \sqrt{2h\log(1/h)},
\end{equation}
for all $0<h<h_0$ and all $0\leq t\leq 1-h$,
and $\sqrt{2}$ is the best constant for which this is true.
\item  $P_W$-almost every $B\in C[0,1]$ is locally H\"{o}lder continuous with index $0<\al<1/2$.
 \item $P_W$-almost every $B\in C[0,1]$ is not differentiable at any $t\in[0,1]$.
  \end{enumerate}
   \end{Theorem}
   See e.g.\ M\"{o}rters \& Peres (2010), \S1.2 and \S 1.3. A path $f\in C[0,1]$ is locally H\"{o}lder continuous with index $\al>0$ if there is $\varep>0$ such that if $|s-t|<\varep$, then $|f(s)-f(t)|\leq C|s-t|^{\al}$ for some $C>0$.
This implies the same property for any $0<\al'<\al$. The value $\al<1/2$ is optimal:  $P_W$-almost every  $B\in C[0,1]$ fails to be locally H\"{o}lder continuous with index $\al>1/2$ (M\"{o}rters \& Peres, 2010,  Remark 1.21). Concerning the critical value $\al=1/2$, the best one can say is that $P_W$-almost every $B$ satisfies  $\inf_{t\in[0,1]}\lim\sup_{h\raw 0} (|B(t+h)-B(t)|/\sqrt{h})=1$ (\emph{ibid.}, Theorem 10.30).

     \begin{Theorem}\label{BM2}
Theorem \ref{BM1} holds \emph{verbatim} (even without any computability assumption on $t\in[0,1]$!) if `for $P_W$-almost every $B\in C[0,1]$' is replaced by `for every $P_W$-random $B\in C[0,1]$'. 
      \end{Theorem}  
\noindent        For continuity see Fouch\'{e} (2008), \S 3, and Allen,  Bienvenu, \& Slaman (2014), \S 2.3.  For non-differentiability see  Fouch\'{e} (2008), Theorem 7. See also   Fouch\'{e} \& Mukeru (2022).
      \section{Applications to statistical mechanics}\label{SM}
            \begin{quote}
\begin{small}
It is the author's view that many of the most important questions still remain unanswered in very fundamental and important ways. (Sklar, 1993, p.\ 413)\end{small}
\end{quote}
\begin{quote}
\begin{small}
What many ``chefs''  regard as absolutely essential and indispensable, is argued to be insufficient or superfluous by many others. (Uffink, 2007, p. 925)\end{small}
\end{quote}
The theme of the previous section is the mathematical key to a physical understanding of the notorious phenomenon of \emph{irreversibility},  for the moment in classical statistical mechanics. The literature on this topic is enormous; I recommend Sklar (1993), Uffink (2007), and Bricmont (2022). My discussion is based on the pioneering work of Hiura \& Sasa (2019). But before getting there, I would like to very briefly review Boltzmann's take on the general problem of irreversibility (which in my view is correct).
In Boltzmann's approach,  irreversibility of macroscopic phenomena in a microscopic world governed by Newton's (time) reversible equations is a consequence of:
\begin{enumerate}
\item \emph{Coarse-graining} (only certain macroscopic quantities behave irreversibly);
\item \emph{Probability} (irreversible behaviour is just very likely--or,  in infinite systems, almost sure).
\end{enumerate}
Boltzmann launched two different scenarios to make this work, both extremely influential. First, in Boltzmann (1872) the coarse-graining
 of an $N$-particle system moving in some volume $V\subset \R^3$ was done by introducing a time-dependent macroscopic distribution function $f_t$, which for each time $t$ is defined on the single-particle phase space $V\x\R^3\subset \R^6$, and which is a probability density in the sense that $N\cdot \int_A d^3\mathbf{r}d^3\mathbf{v}\, f_t(\mathbf{r},\mathbf{v})$  is the ``average'' number of particles inside a region $A\subset V\x\R^3$ at time $t$ (the normalization is $\int d^3\mathbf{r}d^3\mathbf{v}\, f_t(\mathbf{r},\mathbf{v})=1$).  Boltzmann  argued that under various assumptions, notably his \emph{Stosszahlansatz} (``molecular chaos'') that assumes probabilistic independence of two particles before they collide, 
 as well as the absence of collisions between three or more particles, and finally some form of smoothness, $f_t^{(N)}$
  solves the  \emph{Boltzmann equation} 
 \beq
 \partial_t f_t +\mathbf{v}\cdot \partial_{\mathbf{r}}f_t = C\left(f_t\right), \label{BEq}
 \eeq whose right-hand side is a quadratic integral expression in $f$ taking the effect of two-body-collisions (or other two-particle interactions) into account. He then showed that
 the ``entropy''  
  \beq
  S(t) = - \int_{\R^6} d\mathbf{r}d\mathbf{v}\,  f_t(\mathbf{r}, \mathbf{v},t) \ln f_t(\mathbf{r}, \mathbf{v},t)
  \eeq
satisfies $dS/dt\geq 0$  whenever $f_t$ solves his equation, and saw this as a 
 proof of irreversibility.
 
Historically, there were two immediate objections to this result (see also the references above). First,
 there
 is some tension between this irreversibility and the reversibility of Newton's equations satisfied by the microscopic variables $(\mathbf{r}_0(t), \mathbf{v}_0(t), \ldots  \mathbf{r}_{N-1}(t),\mathbf{v}_{N-1}(t))$ on which 
 $f_t$ is based (Loschmidt's \emph{Umkehreinwand}). Second,   in a finite system any $N$-particle configuration eventually returns to a configuration arbitrarily close to its initial value (Zermelo's \emph{Wiederkehreinwand}). 
 A general form of this phenomenon of \emph{Poincar\'{e} recurrence} (e.g.\ Viana \& Oliveira, 2016, Theorem 1.2.1) states that if $(X,P,T)$ is a dynamical system, where  $P$ is $T$-invariant and $T:X\raw X$ is just assumed to be measurable, and $A\subset X$ has positive measure, then for $P$-almost every $x\in E$ there exists infinitely many  $n\in\N$ for which $T^n(x)\in A$. 
 These problems made Boltzmann's conclusions look dubious, perhaps even circular  (irreversibility having been put in by hand via the assumptions leading to the Boltzmann equation).
Despite the famous later work by Lanford (1975, 1976) on the derivation of the Boltzmann equation for short times, these issues  remain controversial, see e.g.\  the debate between Uffink \& Valente (2015) and Ardourel (2017). But I see a  promising way forward, as follows (Villani, 2002, 2013; Bouchet, 2020; Bodineau \emph{at al.}, 2020). From the point of view of Boltzmann (1877), as rephrased in \S\ref{BEP} (first bullet), the distribution function is just the empirical measure \er{LN} for an $N$-particle system with $A=\R^6$ (hence $A$ is uncountably infinite, but nonetheless the measure-theoretic situation remains unproblematic). Each $N$-particle configuration 
 \beq
 x^{(N)}(t):=(\mathbf{r}_0(t), \mathbf{v}_0(t), \ldots  \mathbf{r}_{N-1}(t),\mathbf{v}_{N-1}(t))\label{64}
 \eeq
  at time $t$ determines a probability measure $L_N( x^{(N)}(t))$ on $A$ via
 \begin{equation}
L_N( x^{(N)}(t))=\frac{1}{N}\sum_{n=0}^{N-1} \dl_{(\mathbf{r}_n(t) \mathbf{v}_n(t))}.\label{LNB}
\end{equation}
Physicists  prefer densities (with respect to Lebesque measure $d\mathbf{r}d\mathbf{v}$)  and Dirac $\dl$-functions, writing
 \begin{align}
f ^{(N)}_t: A^N&\raw  \mathrm{Dis}(A); \\
f_t^{(N)}(\mathbf{r}_0, \mathbf{v}_0, \ldots  \mathbf{r}_{N-1},\mathbf{v}_{N-1}): (\mathbf{r}, \mathbf{v})
&\mapsto
 \frac{1}{N}\sum_{k=0}^{N-1} \dl(\mathbf{r} - \mathbf{r}_k(t)) \dl(\mathbf{v} - \mathbf{v}_k(t)), \label{deff}
 \end{align}
 where $\mathrm{Dis}(A)$ is the space of probability \emph{distributions} on $A$. The connection is  (Villani, 2010, \S1.3) 
 \begin{equation}
dL_N(x^{(N)}(t))=f_t^{(N)}(\mathbf{r}_0, \mathbf{v}_0, \ldots  \mathbf{r}_{N-1},\mathbf{v}_{N-1})\, d\mathbf{r}d\mathbf{v}.\label{68}
\end{equation}
The hope, then, is that $f_t^{(N)}$ has a limit $f_t$  as $N\raw\infty$
that has some smoothness and satisfies the Boltzmann equation.  To accomplish this, the idea is to turn $(f_t^{(N)})_{t\geq 0}$ into a stochastic process taking values in $\mathrm{Dis}(A)$ or in $\mathrm{Prob}(A)$, based on a probability space
$(A^{\om},P^{\om})$, indexed by $N\in\N$, and study the limit $N\raw\infty$. 
More precisely, in a \emph{dilute gas} (for which the Boltzmann equation was designed) one has 
  $a^3\ll  1/\rho \ll \ell^3$, where $a$ is the  atom size (or some other  length scale), $\rho= N/V$ is the particle density, and $\ell$ is the mean free path (between collisions). Defining
   $\varep= 1/(\rho\ell^3)$,  the limit ``$N\raw \infty$'' is the  \emph{Boltzmann--Grad limit} $N\raw\infty$ and 
    $\varep \raw 0$ at constant $\varep N$.
    
The simplest way to put probability measures on  $A^N$ and $A^{\om}$  is  to start from some initial value $f_0$, which 
 is the density of a probability measure $p$ on $A^N$, and hence defines a Bernoulli
probability measure $P_p^{\om}$ on $X=A^{\om}$.   There are two ways to block correlations   in the spirit of the \emph{Stosszahlansatz}. One is to
take permutation-invariant probability measures $(P^{(N)})$ on $A^N$
for which the empirical measures $L_N$ on $A$ converge (in law) to some $p\in\mathrm{Prob}(A)$ as $N\raw\infty$ (Villani, 2013); this is equivalent to the factorization $\lim_{N\raw\infty} \la g_1\ot g_2\ot 1_A\cdots \ot 1_A\ra _{P^{(N)}}=\la g_1\ra_{\mu}\la g_2\ra_{\mu}$ for all $g_1, g_2\in C_b(A)$, cf.\ Sznitman (1991), Prop.\ 2.2. Alternatively,
 take $\mu\in\mathrm{Prob}(A)$ and average Bernoulli measures $P_p^{\om}$ with respect to $\mu$, as in de Finetti's theorem (cf.\ Aldous, 1985). 
Either way, one's hope is that \emph{$P^{\om}$-almost surely} the random variables $f_t^{(N)}$ have a smooth limit $f_t$ as $N\raw\infty$, which limit distribution function satisfies the Boltzmann equation, so that the macroscopic time-evolution $t\mapsto f_t$ is induced by the microscopic time-evolution $t\mapsto x(t)$  at least for the  $P^{\om}$-a.e.\ $x\in X$ for which $\lim_{N\raw \infty} f_t(x)$ exists, where $x$ is some configuration of infinitely many particles in $\R^3$, including  their velocities, cf.\ \er{64} - \er{68}.
 This would  even \emph{derive} the Boltzmann equation.
 
  Using large deviation theory, Bouchet (2020) showed all this \emph{assuming} the \emph{Stosszahlansatz}. This is very impressive, but the argument would be complete only if one could prove that, in the spirit of the previous section, $\lim_{N\raw \infty} f^{(N)}_t(x)$ exists \emph{for all $P^{\om}$-random} $x\in X$, preferably by showing first that the  \emph{Stosszahlansatz} and the other assumptions used in the derivation of the  Boltzmann equation (such as the absence of multiple-particle collisions) hold for all $P^{\om}$-random $x$.  In particular, this would make it clear that the  \emph{Stosszahlansatz} is really a randomness assumption.
  Earman's prinicple applies: Bouchet (2020) showed that for finite $N$, the Boltzmann equation holds approximately for a set of initial conditions $x\in A^N$ with high probability $P^N$. The resolution of the \emph{Umkehreinwand} is then standard, see e.g.\ Bricmont (2022), Chapter 8. Similarly, the \emph{Wiederkehreinwand} is countered by noting that in an infinite system the recurrence time is infinite, whilst in a large system it is  astronomically large (beyond the age of the universe).

While its realization for  the Boltzmann equation may still be remote (for mathematical rather than conceptual reasons or matters of principle), this scenario can be demonstrated in the \emph{Kac ring model} (Hiura \& Sasa, 2019). The original reference for the Kac ring model is Kac (1959); useful literature prior to the use of algorithmic randomness in Hiura \& Sasa (2019) includes 
Maes, Neto\u{c}n\'{y}, \& Shergelashvili (2007) and  De Bi\`{e}vre \&  Parris (2017). 
 It is a caricature of the Boltzmann equation rather in the spirit of Boltzmann (1877), i.e.\ his second approach to the problem of irreversibility (the state counting techniques reviewed in \S\ref{BEP} come from this second paper). Namely:
 \begin{itemize}
\item The \emph{microstates} of the Kac ring model for finite $N$ are pairs 
 \begin{align}
(x^{(N)}, y^{(N)})\in
2^{2N+1}\x 2^{2N+1}\equiv A^N; &&  x^{(N)}=(x_{-N}, \ldots, x_N); &&  y^{(N)}=(y_{-N}, \ldots, y_N), 
\end{align}
with $x_n\in 2,\, y_n\in 2$. Here $x_n$ is seen as a spin that can be ``up'' ($x_n=1$) or ``down'' ($x_n=0$), whereas $y_n$ denotes the presence ($y_n=1$) or absence ($y_n=0$) of a  
scatterer,  located between $x_n$ and $x_{n+1}$. 
These replace the variables 
 $(\mathbf{r}_0, \mathbf{v}_0, \ldots,  \mathbf{r}_{N-1},\mathbf{v}_{N-1})\in\R^{6N}$ for the Boltzmann equation.
In the thermodynamic limit we then have $(x^{(N)}, y^{(N)})\stackrel{N\raw\infty}{\longrightarrow} (x,y)\in 2^{\Z}\x 
  2^{\Z}\equiv A^{\om}$.
  \item The \emph{macrostates} of the model, which replace the distribution function \er{deff}, form a pair
 \begin{align}
  m^{(N)}: A^N\raw [0,1], && m^{(N)}(x^{(N)}, y^{(N)}):=\frac{1}{2N+1}\sum_{k=-N}^N x_k; 
  \\
  s^{(N)}: A^N\raw [0,1], &&   s^{(N)}(x^{(N)}, y^{(N)})=\frac{1}{2N+1}\sum_{k=-N}^N y_k.
\end{align}
  \item The \emph{microdynamics} replacing the time evolution $(\mathbf{r}_0(t), \mathbf{v}_0(t), \ldots,  \mathbf{r}_{N-1}(t),\mathbf{v}_{N-1}(t))$ generated by Newton's equations with some  potential, is now discretized, and is given by maps
\begin{align}
T^{(N)}: A^N\raw A^N; &&
   T^{(N)} (x,y)_{n+1}& :=(x_n, y_n) && (y_n=0);\\
&&   &:=(1-x_n,y_n) && (y_n=1),\label{MID}
\end{align}
  where $n=-N,\ldots, N$, with periodic boundary conditions, i.e.\ $(x_{N+1},y_{N+1})=(x_{-N}, y_{-N})$.
  The same formulae define the thermodynamic limit $T:A^{\om}\raw A^{\om}$.
 The idea is that in one time step the spin $x_n$ moves one place to the right and flips iff it hits a scatterer ($y_n=1$).
  \item The \emph{macrodynamics}, which replaces the solution of the Boltzmann equation, is given by
  \begin{align}
  \Phi: [0,1]\x [0,1]\raw  [0,1]\x [0,1]; &&   \Phi(\ovl{m},\ovl{s})=
  ((1-2\ovl{s})(\ovl{m}-\half)+\half,\ovl{s});\label{MAD} 
    \end{align}
    In particular, for $t\in\N$  one has 
    \begin{equation}
    \Phi^t(\ovl{m},\ovl{s})=
  ((1-2\ovl{s})^t(\ovl{m}-\half)+\half,\ovl{s})  , \label{BEKac}
    \end{equation} and hence every initial state $(\ovl{m},\ovl{s})$ with  $\ovl{s}\in(0,1)$ reaches the ``equilibrium'' state $(\half,\ovl{s})$, as
    \begin{equation}
\lim_{t\raw\infty}     \Phi^t(\ovl{m},\ovl{s})=(\half,\ovl{s}).
\end{equation}
  \item The macrodynamics \er{MAD} is induced by the microdynamics \er{MID}, that is,
  \begin{equation}
(m^{(N)}, s^{(N)})(T^{(N)}(x^{(N)}, y^{(N)}))=\Phi((m^{(N)}, s^{(N)})(x^{(N)}, y^{(N)})),
\end{equation}
provided the counterpart of the
    \emph{Stosszahlansatz} for this model holds. For $N<\infty$ this reads  
    \beq
    \#_{x=1}(t+1)=(1-\ovl{s})\,\#_{x=1}(t)+ \ovl{s}\,\#_{x=0}(t), \label{SZA}
    \eeq
i.e., the number of spins $x=1$ after one time step equals the number of spins that already had $x=1$ and did not scatter (where the probability of non-scattering  is estimated to be $1-\ovl{s}$, i.e., 
  the average density of voids), plus the number of spins
 $x=0$ that have flipped because they hit a scatterer (where the probability of scattering is estimated to be the average density $\ovl{s}$ of scatterers). This kind of averaging of course overlooks the details of the actual location of the scatterers versus the location of the spins with specific values. It is trivial to find configurations $(x,y)$ where it is violated, but these become increasingly rare as $N\raw\infty$.
\end{itemize}
I now state Theorem 3.5 in Hiura \& Sasa (2019), which sharpens earlier results by Kac (1959) in replacing a `for $P$-almost every $x$' result by a `for all $P$-random $x$' result that provides much more precise information on randomness.
First, recall that if   $(x,y)\in A^{\om}$ is
$P^{\om}_{\ovl{m}}\x P^{\om}_{\ovl{s}}$-random, then
\begin{equation}
\lim_{N\raw\infty} (m^{(N)}, s^{(N)})(x^{(N)}, y^{(N)})=(\ovl{m},\ovl{s}).
\end{equation}\vspace{-5mm}
\begin{Theorem}\label{HS} 
For all computable macrostates $(\ovl{m},\ovl{s})\in [0,1]\x [0,1]$  and all $P^{\om}_{\ovl{m}}\x P^{\om}_{\ovl{s}}$-random microstates $(x,y)\in A^{\om}$, the macrodynamics \er{MAD} is  induced by the microdynamics \er{MID} as $N\raw\infty$:
\begin{equation}
\lim_{N\raw\infty}(m^{(N)}, s^{(N)})(T^{(N)}(x^{(N)}, y^{(N)}))=\Phi(\ovl{m},\ovl{s}).
\end{equation}
\end{Theorem}
\noindent It follows that the ``Boltzmann equation'' \er{BEKac} holds, and that the macrodynamics is \emph{autonomous}: the dynamics of the macrostates $(\ovl{m},\ovl{s})$ does not \emph{explicitly} depend on  the underlying microstates. 

Theorem \ref{HS}  uses \emph{biased} Martin-L\"{o}f randomness on $A^{\om}$ and hence defines ``typicality'' \emph{out of equilibrium}. As we have seen, equilibrium corresponds to $\ovl{m}=\half$ (for arbitrary $\ovl{s}\in(0,1)$), for which the corresponding $P^{\om}_{1/2}$-random states are arguably the most random ones: for it follows from  \er{EPLG} that if $s$ is $P^{\om}_p$-random for some $p\in (0,1)$ and $s'$ is $P^{\om}_f$-random, then 
\beq
K(s_{|N}) \leq K(s'_{|N}),
\eeq
so that the  approach to equilibrium $\ovl{m}\raw 1/2$ increases (algorithmic)  randomness, as expected.

 The  same perspective arises from entropy  (Maes, Neto\u{c}n\'{y}, \& Shergelashvili, 2007; Hiura \& Sasa, 2019). The \emph{fine-grained (microscopic)} entropy of some $P^{\om}\in\mathrm{Prob}(A^{\om})$, may be defined by 
\begin{equation}
h(P^{\om}):= \lim\sup_{N\raw\infty} -\frac{1}{N} 
\sum_{(x^{(N)}, y^{(N)})\in A^N} P^{\om}([x^{(N)}, y^{(N)}])
\ln P^{\om}([x^{(N)}, y^{(N)}]).
\end{equation} For example, 
as in \er{NH2}, the Bernoulli measure $P^{\om}=P^{\om}_{(\ovl{m},\ovl{s})}$ has fine-grained entropy
\begin{align}
h(P^{\om}_{(\ovl{m},\ovl{s})})=h_2(\ovl{m},\ovl{s})=h_2(\ovl{m})+ h_2(\ovl{s}); &&h_2(\ovl{m}) =
-\ovl{m}\log_2 \ovl{m} -(1-\ovl{m})\log_2 (1-\ovl{m}),\label{KRMA} 
\end{align}
on which \er{folklore}  gives an algorithmic perspective: for all $P^{\om}_{(\ovl{m},\ovl{s})}$-random microstates $(x,y)$ we have
\beq
 h(P^{\om}_{(\ovl{m},\ovl{s})})= \lim\sup_{N\raw\infty}\frac{K((x,y)_{|N})}{N}.
 \eeq
On the other hand, the \emph{coarse-grained (macroscopic)}  entropy \er{BE} for the flat prior $p=f$ on $A$ and the  probability $\mu=\mu_{(\ovl{m},\ovl{s})}$  on $2\x 2$ defined by  $\mu_{(\ovl{m},\ovl{s})}(1,0)=\ovl{m}\cdot (1-\ovl{s})$ etc.\ is given by
\beq
s_B(\mu(\ovl{m},\ovl{s}))=h_2(\ovl{m},\ovl{s})-2\ln 2,  \label{KRMI} 
\eeq 
cf.\  \er{SBflat}.
Despite the  similarity of \er{KRMA} and \er{KRMI}, we should keep them apart.
 Irreversibility of the macroscopic dynamics does not contradict reversibility of the microscopic dynamics, even though the fine-grained and coarse-grained entropies practically coincide here. In this case, defining time reversal $\ta:A^{\om}\raw A^{\om}$ by 
 \beq
 \ta(x,y)_n:=(x_{-n},y_{-n-1}),
 \eeq so that $\ta\circ T^{(N)}= (T^{(N)})\inv\circ \ta$,  one even has $\Phi\circ\ta=\Phi$. Here $(T^{(N)})\inv$ is given by \er{MID} with $n+1\leadsto n-1$ (where $\leadsto$ stands for `is replaced by') and $y_n=0/1\leadsto y_{n-1}=0/1$, so that the spin now moves to the left. But the real point is that if $(x,y)$ is $P^{\om}_{(\ovl{m},\ovl{s})}$-random, then ``typically'', $\ta(x,y)$ is not.
In fact, the entire (neo) Boltzmannian program can be carried out in this model (Maes, Neto\u{c}n\'{y}, \& Shergelashvili, 2007;  Bricmont, 2022, \S 8.7.2; Hiura \& Sasa, 2019).
 In particular, 
 the coarse-grained entropy \er{KRMA} is invariant under the microscopic time evolution $T$, whereas the fine-grained entropy \er{KRMI} increases along solutions of the ``Boltzmann equation'' \er{BEKac}.
\section{Applications to quantum mechanics}\label{QM}
There is yet another interpretation of the diagram at the beginning of \S\ref{BEP}:
 in \emph{quantum mechanics} a string $\sg\in A^N$ denotes the outcome of a run of $N$ repeated measurements of the same observable  $\mathbf{A}$ with finite spectrum $A$ in the same quantum state, so that the possible outcomes $a\in A$ are distributed according to the Born rule: if $H$ is the Hilbert space pertinent to the experiment, $\mathbf{A}\in B(H)$ is the observable that is being measured, with spectrum $A=\sg(\mathbf{A})$ and spectral projections $\mathbf{E}_a$ onto the eigenspace $H_a$ for eigenvalue $a$,
  and $\hat{\rho}$ is the density operator describing the quantum state, then  $p(a)=\Tr(\hat{\rho}\mathbf{E}_a)$.   
   It can be shown that if we consider the run as a single experiment, the probability of outcome $\sg$ is  $P^{N}_p(\sg)$, as in a classical Bernoulli trial. This  extends to the idealized case of an infinitely repeated experiment, described by the  probability measure $P^{\omega}_p$ on 
  $A^{\omega}$ (Landsman, 2021). In particular, for a ``fair quantum toss'' (in which $A=2$ with $p(1)=p(0)=1/2$),  it follows that the outcome sequences sample the probability space $(2^{\om}, P^{\om}_f)$, just as in the classical case.
  
  For quantum mechanics itself, this implies that $P^{\om}_f$-almost every  outcome sequence $s\in 2^{\om}$ is  $P^{\om}_f$-random.  The theme of \S\ref{PP} then leads to the circular conclusion that
  all  $P^{\om}_f$-random outcome sequences are  $P^{\om}_f$-random. Nonetheless, this circularity comes back with a vengeance if we turn to hidden variable theories, notably
    Bohmian mechanics  (cf.\ Goldstein, 2017). Let me first summarize my original argument (Landsman, 2021, 2022), and then reply to a potential counterargument.
    
In \hv\ theories there is a space $\Lm$ of hidden variables, and if the theory has the right to call itself ``deterministic'', then  there must be functions $h:\N\raw\Lm$ and $g:\Lm\raw A$ such that 
  \beq
  s=g\circ h.\label{sgh}
  \eeq 
  The existence of $g$ expresses the idea that  the value of $\lm$ \emph{determines} the outcome of the experiment.
  The function $g$ tacitly incorporates all details of the experiment that may affect its outcome,   \emph{except the hidden variable} $\lm$ (which is the argument of $g$). Such details may include
   the setting,  a possible context, and the quantum state. The existence of $g$ therefore does not contradict the Kochen--Specker theorem (which excludes context-dependence). But $g$ is just \emph{one}  ingredient that makes  a hidden variable theory deterministic. The other is the function $h$ that gives the value of $\lm$ in experiment No.\ $n$ in a long run, for each $n$. Furthermore, in any hidden variable theory the probability of the outcome of some measurement \emph{if the \hv\ $\lm$ is unknown} is given by averaging the determined outcomes given by $g$ with respect to some probability measure $\mu_{\psi}$  on  $\Lm$ defined by the quantum state $\psi$ supposed to describe the experiment within \qm.
   \begin{Theorem}\label{TAB}
 The  functions $g$ and  $h$ cannot \emph{both} be provided by any deterministic theory (and hence deterministic \hv\ theories that exactly reproduce the Born rule cannot exist). 
   \end{Theorem}
   
\emph{Proof.}  The Born rule is needed to prove that outcome sequences $s\in A^{\om}$ are $P^{\om}_p$-distributed (Landsman, 2021, Theorem 3.4.1).
If $g$ and $h$ were explicitly given by some deterministic theory $T$, then the sequence $s$ would be described explicitly via \er{sgh}. By (what I call) \emph{Chaitin's second incompleteness theorem},
 the sequence $s$ cannot then be $P^{\om}_p$-random.
\hfill Q.E.D.
 \smallskip
 
 The theorem used here states that if $s\in A^{\om}$ is $P^{\om}_p$-random, 
then ZFC (or any  sufficiently comprehensive mathematical theory $T$ meant in the proof of Theorem \ref{TAB}) can compute only finite many digits of $s$.  See e.g.\ Calude (2002), Theorem 8.7, which is stated for Chaitin's famous random number $\Omega$ but whose proof holds for any $P^{\om}_p$-random sequence. Consistent with Earman's principle, Theorem \ref{TAB} does not rely on the idealization of infinitely long runs of measurements, since for finite runs \emph{Chaitin's (first) incompleteness theorem} leads to a similar contradiction. The latter theorem states that for any sound mathematical theory $T$ containing enough arithmetic there is a constant $C\in\N$ such that $T$ cannot prove any sentence of the form $K(\sg)>C$ \emph{although infinitely many such sentences are true}. In other words, $T$ can only prove  randomness of finitely many strings, \emph{although infinitely many strings are in fact random}.  See e.g.\ Calude (2002), Theorem 8.4.
   
  The upshot is that a deterministic theory cannot produce random sequences. Against this, fans of deterministic \hv\ theories could argue that the (unilateral) Bernoulli shift $S$ on $2^{\om}$ (equipped with 
  $P^{\om}_f$ for simplicity) is deterministic and yet is able to produce random sequences.
  
 Indeed, following a suggestion by Jos Uffink (who is not even an Bohmian!),
  this can be done  as follows;  readers familiar with D\"{u}rr, Goldstein, \& Zanghi (1992) will notice that the scenario in the main text would actually be optimal for these authors. With $\Lm=A=2$, and the simplest experiment for which $g:2\raw 2$ is  the identity (so that the measurement just reveals the actual value of the pertinent \hv), 
   take an initial condition  $s'\in 2^{\om}$, and define
    $h:\N\raw 2$  by 
    \beq
    h(n)=s'(n).
    \eeq Then 
$s=s'$. 
    In other words, imagine that
  experiment number $n\in\N$  takes place at time $t=n$, at which time the \hv\ takes the value $\lm=s'(n)$. The 
  measurement run then just reads the tape $s'$. Trivially, if the initial condition $s'$ is $P^{\om}_p$-random, then  so is the outcome sequence $s$.
   
 According to D\"{u}rr, Goldstein, \& Zanghi (1992),
 the  randomness of outcomes in the deterministic world  envisaged in Bohmian mechanics originates in the random initial condition of universe, which is postulated to be in ``quantum equilibrium''. In the above toy example, the configuration space (which in Bohmian mechanics is $\R^{3N}$) is replaced and idealized by $2^{\om}$, i.e.\ the role of the position variable $q\in \R^{3N}$ is now played by $s\in 2^{\om}$;
 the dynamics (replacing the Schr\"{o}dinger equation) is $S$; and  the ``quantum equilibrium condition'' (which is nothing but the Born rule) then postulates that its initial value $s'$ is distributed according to  the Born rule, which here is the fair Bernoulli measure $P^{\om}_f$.
 The Bohmian explanation of randomness then comes down to the claim that despite the determinism inherent  in the dynamics $S$ as well as in the measurement theory $g$:
 \begin{center}
  \emph{Each experimental outcome $s(n)$ 
is random because the hidden variable $\lm$  is randomly distributed}.
Since $s'=s$, this simply says that \emph{$s$ is random because $s$ is random}.
\end{center}
Even in less simplistic scenarios, using the language of computation theory (taking computability as a metaphor for determinism) we may say: deterministic \hv\ theories need a random oracle to reproduce the randomness required for \qm. This defeats their determinism. 
\section{Summary} 
This paper was motivated by a number of (closely related) questions, including these:
\begin{enumerate}
\item  Is it probability or randomness that ``comes first''? How are these concepts related?
\item Could the notion of ``typicality'' as it is used in Boltzmann-style statistical mechanics (e.g.\ Bricmont, 2022)  be replaced by some precise mathematical form of randomness? 
\item Are ``typical'' trajectories in ``chaotic'' dynamical systems (i.e.\ those with high Kolmogorov--Sinai entropy)  random in the same, or  some similar sense?
\end{enumerate}
Here  ``typical''  means ``extremely probable'',  which may be idealized to ``occurring almost surely''.  My attempts to address these questions are guided by what I call \emph{Earman's principle}, stated after Theorem \ref{Calthm} in \S\ref{AR}, which regulates the connection between actual and idealized physical theories. On this score, $P$-randomness (see \S\ref{AR})  does quite well, cf.\ Theorems \ref{T3.4} and \ref{Calthm}, although I have some misgivings about the physical relevance of its mathematical origins in the theory of computation, which for physical applications should be replaced by some abstract logical form of determinism. 

Various mathematical examples provide situations where some property $\Phi(x)$ that holds for $P$-almost every $x\in X$ (where $P$ is some probability measure on $X$) in fact holds for all $P$-random $x\in X$, at least under some further computability assumptions, see \S\ref{PP}.  The main result in \S\ref{SM}, i.e.\ Theorem \ref{HS} due to Hiura \& Sasa (2019), as well as the much better known results about the relationship between entropy, dynamical systems, and $P$-randomness reviewed \S\ref{BEP} and \S\ref{PP}, notably Theorem \ref{myT1} and eq.\ \er{Brudno}, provide positive answers to  questions 2 and 3. This, in turn, paves the way for an explanation of emergent phenomena like irreversibility and chaos, and suggests that the answer to question 1 is that at least the computational concept of $P$-randomness requires a prior probability $P$!
\addcontentsline{toc}{section}{References}
\begin{small}

\end{small}
\end{document}

%% file: Bohrificationpreamble.tex
\newcommand{\id}[1]{\ensuremath{\mathrm{id}}}

\newcommand{\Q}{\mathbb{Q}}

\newcommand{\half}{\mbox{\footnotesize $\frac{1}{2}$}}

\newcommand{\qm}{quantum mechanics}
\newcommand{\er}{\eqref}
 
\newcommand{\beq}{\begin{equation}}
\newcommand{\eeq}{\end{equation}} 
\newcommand{\bea}{\begin{eqnarray}}
\newcommand{\eea}{\end{eqnarray}}

\newcommand{\ovl}{\overline}

\newcommand{\raw}{\rightarrow}

\newcommand{\ot}{\otimes} 
\newcommand{\la}{\langle} \newcommand{\ra}{\rangle}

\newcommand{\x}{\times}

\newcommand{\hv}{hidden variable}

\newcommand{\al}{\alpha} 
 \newcommand{\Gm}{\Gamma}
\newcommand{\dl}{\delta} \newcommand{\Dl}{\Delta}
 \newcommand{\varep}{\varepsilon}

\newcommand{\lm}{\lambda} \newcommand{\Lm}{\Lambda}
 \newcommand{\sg}{\sigma}
\newcommand{\Sg}{\Sigma} \newcommand{\ta}{\tau} 
 
\newcommand{\ch}{\ch}  
\newcommand{\om}{\omega} \newcommand{\Om}{\Omega}

\newcommand{\inv}{^{-1}}

\newcommand{\Tr}{\mbox{\rm Tr}\,}

\newcommand{\CB}{{\mathcal B}}

\newcommand{\CO}{{\mathcal O}}

\newcommand{\N}{{\mathbb N}} \newcommand{\R}{{\mathbb R}}
 \newcommand{\Z}{{\mathbb Z}}
  \makeatletter
 
\makeatletter
\def\moverlay{\mathpalette\mov@rlay}
\def\mov@rlay#1#2{\leavevmode\vtop{%
   \baselineskip\z@skip \lineskiplimit-\maxdimen
   \ialign{\hfil$\m@th#1##$\hfil\cr#2\crcr}}}
\newcommand{\charfusion}[3][\mathord]{
    #1{\ifx#1\mathop\vphantom{#2}\fi
        \mathpalette\mov@rlay{#2\cr#3}
      }
    \ifx#1\mathop\expandafter\displaylimits\fi}
\makeatother